\documentclass{article}
\usepackage{amssymb}
\usepackage{graphicx}
\usepackage{amsmath}
\usepackage{amsfonts}

\begin{document}

\title{On Hirschman and log-Sobolev inequalities in \\${\mu}$-deformed Segal-Bargmann analysis}
\author{Claudio de Jes\'{u}s Pita Ruiz Velasco \thanks{Research partially supported by
grant 165628, CONACyT (Mexico)}\\Universidad Panamericana\\Mexico City, Mexico\\email: cpita@mx.up.mx \setcounter{footnote}{6}
\and Stephen Bruce Sontz \thanks{Research partially supported by grants 32146-E and
P-42227-F, CONACyT (Mexico)}\\Centro de Investigaci\'{o}n en Matem\'{a}ticas, A.C. (CIMAT)\\Guanajuato, Mexico\\email: sontz@cimat.mx}
\date{June, 2005}
\maketitle

\begin{abstract}
\noindent We consider a ${\mu}$-deformation of the Segal-Bargmann transform,
which is a unitary map from a ${\mu}$-deformed ground state representation
onto a ${\mu}$-deformed Segal-Bargmann space. We study the ${\mu}$-deformed
Segal-Bargmann transform as an operator between $L^{p}$ spaces and then we
obtain sufficient conditions on the Lebesgue indices for this operator to be
bounded. A family of Hirschman inequalities involving the Shannon entropies of
a function and of its ${\mu}$-deformed Segal-Bargmann transform are proved. We
also prove a parametrized family of log-Sobolev inequalities, in which a new
quantity that we call \textquotedblleft dilation energy\textquotedblright
appears. This quantity generalizes the \textquotedblleft energy
term\textquotedblright that has appeared in a previous work.

\end{abstract}
\tableofcontents

\section{Introduction}

\bigskip The Segal-Bargmann space $\mathcal{B}^{2}$ is the Hilbert space of
holomorphic functions $f:\mathbb{C\rightarrow C}$ which are square integrable
with respect to a Gaussian measure $d\nu_{\text{Gauss}}$. As one thinks of the
Hilbert space $L^{2}\left(  \mathbb{R},dx\right)  $ as a quantum configuration
space, one thinks of $\mathcal{B}^{2}$ as a quantum phase space, since the
spaces $\mathbb{R}$ and $\mathbb{C}$ are the configuration space $\mathbb{R}$
and phase space $T^{\ast}\mathbb{R=R}^{2}\cong\mathbb{C} $ for a classical
particle with one degree of freedom. In each of the quantum spaces
$L^{2}\left(  \mathbb{R},dx\right)  $ and $\mathcal{B}^{2}$ one has unbounded
operators $a^{\ast}$ (creation) and $a$ (annihilation), which satisfy the
relation $\left[  a,a^{\ast}\right]  =I$, called the Canonical Commutation
Relation (CCR), and both Hilbert spaces carry irreducible representations of
the Lie group generated by the exponentiated form of the CCR. The Stone-von
Neumann theorem says that in such a case there exists an essentially unique
unitary operator $\widetilde{B}:L^{2}\left(  \mathbb{R},dx\right)
\rightarrow\mathcal{B}^{2}$ that intertwines the action of the corresponding
creation and annihilation operators. This isomorphism $\widetilde{B}$ is the
\textit{Bargmann transform}, and \textit{Segal-Bargmann analysis} has to do
mainly with the study of the operators related to the transform $\widetilde
{B}$ and spaces of holomorphic functions related to $\mathcal{B}^{2}$. (The
beginnings of this theory date back to the works of Segal [Seg1], [Seg2] and
Bargmann [Bar].) When the quantum configuration space is replaced by another
unitarily equivalent Hilbert space $L^{2}(\mathbb{R},dg)$, called the
\textit{ground state representation} (in which $dg$ is another Gaussian
measure), the resulting transform $B$ that maps the ground state
representation unitarily onto the Segal-Bargmann space is called the
\textit{Segal-Bargmann transform}, and this is the operator that will be of
interest for us in this work. In terms of the operators $a^{\ast}$ and $a$ one
can define operators $P$ (momentum) and $Q$ (position), which are unbounded
self-adjoint operators that satisfy the commutation relation $[P,Q]=-iI$,
which is implied by the CCR. If $H=2^{-1}(Q^{2}+P^{2})$ is the Hamiltonian of
the harmonic oscillator, one has also that the operators $P$ and $Q$ satisfy
the equations of motion\textit{\ }$i[P,H]=Q$ and $i[Q,H]=-P$. In 1950, Wigner
[Wig] proved that the converse implication is false by exhibiting a family of
unbounded operators, labeled by a parameter $\mu>-1/2$, that satisfy the
equations of motion but do not satisfy the CCR. Rosenblum and Marron described
explicitly (in [Ros1], [Ros2] and [Marr]) a $\mu$-quantum configuration space
$L^{2}(\mathbb{R},\left\vert x\right\vert ^{2\mu}dx)$, a $\mu$-Segal-Bargmann
space $\mathcal{B}_{\mu}^{2}$, and a $\mu$-Bargmann transform $B_{\mu}$ which
is a unitary onto transformation mapping the former Hilbert space to the
latter Hilbert space. This theory can be understood as a $\mu$-deformation of
standard Segal-Bargmann analysis with the property that if one sets $\mu=0$
the standard theory is recovered (see [Snt3]).

The Segal-Bargmann transform $B$ shares with the Fourier transform
$\mathcal{F}$ the fact of being a unitary operator between $L^{2}$ spaces.
This is one of the original motivations in [Snt1] for studying $B$ by using
$\mathcal{F}$ as a model. For example, the Fourier transform can be studied as
an operator acting on $L^{p}$ spaces, by looking for pairs of Lebesgue indices
$p$ and $q $ for which $\mathcal{F}:L^{p}\left(  \mathbb{R},dx\right)
\rightarrow L^{q}\left(  \mathbb{R},dx\right)  $ is a bounded operator. The
Hausdorff-Young inequality tells us that for $p\in\left[  1,2\right]  $ and
$q=p^{\prime}$ (the conjugate index of $p$), the operator $\mathcal{F}$ is
bounded and that $\left\Vert \mathcal{F}f\right\Vert _{L^{q}\left(
\mathbb{R},dx\right)  }\leq\left\Vert f\right\Vert _{L^{p}\left(
\mathbb{R},dx\right)  } $. In [Snt1] it is proved that for $1\leq q<2$ and
$p>1+q/2$, the Segal-Bargmann transform $B$ is a bounded operator from
$L^{p}\left(  \mathbb{R},dg\right)  $ to $L^{q}\left(  \mathbb{C}%
,d\nu_{\text{Gauss}}\right)  $. By using the Riesz-Thorin interpolation
theorem it is also proved that if $p$ and $q$ are as before, then one has the
estimate $\left\Vert Bf\right\Vert _{L^{q\left(  s\right)  }\left(
\mathbb{C},d\nu_{\text{Gauss}}\right)  }\leq C^{s}\left\Vert f\right\Vert
_{L^{p\left(  s\right)  }\left(  \mathbb{R},dg\right)  } $, where $((p(s))
^{-1},(q(s)) ^{-1}) $ is a point in the segment connecting $( 1/2,1/2) $ and
$( p^{-1},q^{-1}) $. Observe that this result is in fact a family a
Hausdorff-Young type inequalities (with $B$ replacing $\mathcal{F}$). In
[Hir], Hirschman proved the inequality $S_{L^{2}\left(  \mathbb{R},dx\right)
}\left(  f\right)  +S_{L^{2}\left(  \mathbb{R},dx\right)  }\left(
\mathcal{F}f\right)  \leq0$ where $S_{L^{2}\left(  \mathbb{R},dx\right)
}\left(  \varphi\right)  $ is the entropy (defined in the next section) of the
function $\varphi\in L^{2}\left(  \mathbb{R},dx\right)  $. Following [Hir], in
[Snt1] the second author proved the \textquotedblleft Hirschman
inequality\textquotedblright%

\begin{equation}
C_{1}S_{L^{2}\left(  \mathbb{R},dg\right)  }\left(  f\right)  \leq
C_{2}S_{L^{2}\left(  \mathbb{C},d\nu_{\text{Gauss}}\right)  }\left(
Bf\right)  +C_{3}\left\Vert f\right\Vert _{L^{2}\left(  \mathbb{R},dg\right)
}^{2}.\tag{1.1}%
\end{equation}

The importance of this inequality is that it constrains the values of the
entropy of a function and of its Segal-Bargmann transform. That is, even
though the operator $B$ does not preserve entropy (also proved in [Snt1]), the
inequality above shows that the two entropies can not have arbitrary values.
At this point we mention that from the point of view of the Hilbert space
structure, the ground state representation $L^{2}\left(  \mathbb{R},dg\right)
$ is indistinguishable from the Segal-Bargmann space $\mathcal{B}^{2}$, since
$B$ is a Hilbert space isomorphism. In the case of the Fourier transform, the
famous Heisenberg uncertainty principle tells us that the variance of a
function $f$ and the variance of its Fourier transform $\mathcal{F}f$ are
quantities that can not vary arbitrarily. Thus, the inequality (1.1) can be
understood as a kind of uncertainty principle for Segal-Bargmann analysis. The
rest of the work in [Snt1] is about replacing the standard Segal-Bargmann
space by a similar \textquotedblleft weighted\textquotedblright\:space. A
Hausdorff-Young type family of inequalities is proved by using Stein's
interpolation theorem instead of the Riesz-Thorin theorem. Finally, following
the same kind of ideas that lead to the Hirschman inequality, the
logarithmic-Sobolev inequality%

\begin{equation}
C_{1}S_{L^{2}\left(  \mathbb{R},dg\right)  }\left(  f\right)  \leq
C_{2}S_{L^{2}\left(  \mathbb{C},d\nu_{\text{Gauss}}\right)  }\left(
Bf\right)  +C_{3}\left\langle f,Nf\right\rangle _{L^{2}\left(  \mathbb{R}%
,dg\right)  }+C_{4}\left\Vert f\right\Vert _{L^{2}\left(  \mathbb{R}%
,dg\right)  }^{2}\tag{1.2}%
\end{equation}
is shown, where $\left\langle f,Nf\right\rangle _{L^{2}\left(  \mathbb{R}%
,dg\right)  }$ is the quadratic form associated to the energy (or number)
operator $N=a^{\ast}a$ acting in the ground state representation $L^{2}\left(
\mathbb{R},dg\right)  $. Some explanations about why (1.2) is called a
\textquotedblleft log-Sobolev inequality\textquotedblright are presented in
Section 6. The motivation of the present work was whether results similar to
(1.1) and (1.2) are also valid in the context of the $\mu$-deformed theory of
the Segal-Bargmann transform mentioned above. The answers we obtained are
presented here.

We now outline the content of the work. In Section 2 we give the definitions
and some preliminary results that will be used throughout the work. The Banach
spaces introduced in that section, which will be involved in the $\mu
$-deformed Segal-Bargmann spaces considered in the work (introduced in Section
3), are \textquotedblleft weighted\textquotedblright spaces labeled by a
parameter $\lambda>0$. In the case $\mu=0$ considered in [Snt1], a weight $a$
is introduced, and this parameter is related with $\lambda$ by $\lambda=1+a$.
Also, in the case $p=2 $ and $\mu>-1/2$ in [Marr], a weight $\alpha$ is
introduced which can be identified with our parameter $\lambda$. The case in
which $p\geq1$ and $\mu\geq0$, considered in this work, generalizes the case
treated in [Snt1] and the case treated in [Marr] as well.

In Section 3 we introduce the $\mu$-deformed objects (``\textit{generalized}''
objects, in the nomenclature of Rosenblum and Marron) of Segal-Bargmann
analysis with which we will work. So we introduce the $\mu$-deformed ground
state representation $L^{p}\left(  \mathbb{R},dg_{\mu}\right)  $, the
$\lambda$-weighted $\mu$-deformed Segal-Bargmann space $\mathcal{B}%
_{\mu,\lambda}^{q}$, and the $\mu$-deformed Segal-Bargmann transform $B_{\mu}%
$, for which we are interested in values of $p$, $q$ and $\lambda>0$ such that
$B_{\mu}$ is a bounded operator from $L^{p}\left(  \mathbb{R},dg_{\mu}\right)
$ to $\mathcal{B}_{\mu,\lambda}^{q}$.

In Section 4 we show that if the Lebesgue indices $1<p\leq\infty$\textit{,
}$1\leq q<\infty$\ and the weight $\lambda>1/2$ are such that the inequalities
$p>1+q/(2\lambda)$ and $1\leq q<2\lambda$ hold, then the transform $B_{\mu}$
is a bounded operator from $L^{p}\left(  \mathbb{R},dg_{\mu}\right)  $ to
$\mathcal{B}_{\mu,\lambda}^{q}$. Observe that the sufficient conditions for
this result depend on $\lambda$ but not on $\mu$. By setting $\mu=0$ and
$\lambda=1$ we obtain Theorem 3.1 of [Snt1]. The importance of the weight
$\lambda$ in the codomain space is shown by noting that for any $1<p\leq
\infty$\textit{\ }and\textit{\ }$1\leq q<\infty$, the $\mu$-deformed
Segal-Bargmann transform $B_{\mu}$ is always a bounded operator from
$L^{p}\left(  \mathbb{R},dg_{\mu}\right)  $ to $\mathcal{B}_{\mu,\lambda}^{q}$
provided $\lambda$ is large enough, namely $\lambda>\max(q/2 , q/(2(p-1))$.

From [Ros1], [Ros2] and [Marr] we know that the $\mu$-deformed Segal-Bargmann
transform $B_{\mu}$ is a unitary operator from $L^{2}\left(  \mathbb{R}%
,dg_{\mu}\right)  $ onto $\mathcal{B}_{\mu,\lambda}^{2}$, provided $\lambda
=1$. In particular we have that for $p=q=2$ and $\lambda=1$, the operator
$B_{\mu}$ is bounded. We prove in Section 5 that the condition $\lambda=1$ is
also necessary for $B_{\mu}$ to be a unitary operator from $L^{2}\left(
\mathbb{R},dg_{\mu}\right)  $ to $\mathcal{B}_{\mu,\lambda}^{2}$.

The discussion about Hausdorff-Young type inequalities and Hirschman
inequalities is presented in Section 5. In that section we work with the $\mu
$-deformed Segal-Bargmann transform $B_{\mu}$ as an operator from
$L^{p}\left(  \mathbb{R},dg_{\mu}\right)  $ to the unweighted $\mu$-deformed
Segal-Bargmann space $\mathcal{B}_{\mu}^{q}$. We take the parameters $p$ and
$q$ such that the inequalities $p>1+q/2$ and $1\leq q<2$ hold, which imply
that $B_{\mu}$ is a bounded operator. In the case $p=q=2$ the operator
$B_{\mu}$ is also bounded since in this case $B_{\mu}$ is unitary. By applying
the Riesz-Thorin interpolation theorem, we obtain estimates of the operator
norm of $B_{\mu}$ as an operator from $L^{p_{s}}\left(  \mathbb{R},dg_{\mu
}\right)  $ to $\mathcal{B}_{\mu}^{q_{s}}$, $s\in\left[  0,1\right]  $, where
$(p_{s} ^{-1},q_{s}^{-1}) $ is a point in the segment connecting $(
2^{-1},2^{-1}) $ with $(p^{-1},q^{-1}) $. In this way we obtain a
Hausdorff-Young type inequality. This inequality has the property that if we
set $s=0$ in it, the inequality becomes an equality, and this fact plays an
important role in the idea (called the \textquotedblleft differentiation
technique\textquotedblright) in the proof of the Hirschman inequality proved
in this section. The inequality we obtain is%

\begin{equation}
C_{1}S_{L^{2}\left(  \mathbb{R},dg_{\mu}\right)  }\left(  f\right)  \leq
C_{2}S_{L^{2}\left(  \mathbb{C\times Z}_{2},d\nu_{\mu}\right)  }\left(
B_{\mu}f\right)  +C_{3}\left\Vert f\right\Vert _{L^{2}\left(  \mathbb{R}%
,dg_{\mu}\right)  }^{2}.\tag{1.3}%
\end{equation}

If we set $\mu=0$ we recover the inequality (1.1). (The explanation of why the
probability measure space $(\mathbb{C},d\nu_{\text{Gauss}})$ is recovered by
setting $\mu=0$ and $\lambda=1$ in the measure space $\left(  \mathbb{C\times
Z}_{2},d\nu_{\mu,\lambda}\right)  $ is included in Section 2.) Nevertheless we
mention that the proof presented in [Snt1] of the inequality (1.1) works for
all $f\in L^{2}\left(  \mathbb{R},dg\right)  $, while the proof of (1.3)
presented here is only valid for functions $f$ in a dense subspace of
$L^{2}\left(  \mathbb{R},dg_{\mu}\right)  $.

In Section 6 we prove a log-Sobolev inequality following the same steps of the
proof of (1.2) in [Snt1]. That is, by using Stein's interpolation theorem we
prove first a weighted Hausdorff-Young type inequality, and then by applying
the differentiation technique of [Hir] to it, we get the desired log-Sobolev
inequality. In the process of proving (1.2) there appears naturally an energy
term $\left\langle Bf,\widetilde{N}Bf\right\rangle _{\mathcal{B}^{2}}$, which
is the quadratic form associated to the energy operator $\widetilde{N}$ acting
in the Segal-Bargmann space $\mathcal{B}^{2}$ (see [Snt1], pp. 2413-14). But
the unitarity of the Segal-Bargmann transform gives us that the energy
$\left\langle Bf,\widetilde{N}Bf\right\rangle _{\mathcal{B}^{2}}$ is equal to
the energy $\left\langle f,Nf\right\rangle _{L^{2}\left(  \mathbb{R}%
,dg\right)  }$ for $f\in L^{2}\left(  \mathbb{R},dg\right)  $, where
$N=B^{-1}\widetilde{N}B$ is the energy operator acting on the ground state
representation $L^{2}\left(  \mathbb{R},dg\right)  $. It is this latter term
which appears in (1.2). In the $\lambda$-weighted $\mu$-deformed situation we
are dealing with there will appear a new mathematical object that generalizes
the energy term $\left\langle Bf,\widetilde{N}Bf\right\rangle _{\mathcal{B}%
^{2}}$ (corresponding to the Segal-Bargmann transform of $f\in L^{2}\left(
\mathbb{R},dg\right)  $). We will call it \textquotedblleft dilation
energy\textquotedblright\ and denote it by $E_{\mu,\lambda}\left(  B_{\mu
}f\right)  $ (corresponding to the $\mu$-deformed Segal-Bargmann transform of
$f\in L^{2}\left(  \mathbb{R},dg_{\mu}\right)  $). The log-Sobolev inequality
we prove in Section 6 is%

\begin{equation}
C_{1}S_{L^{2}\left(  \mathbb{C\times Z}_{2},d\nu_{\mu}\right)  }\left(
B_{\mu}f\right)  +C_{2}S_{L^{2}\left(  \mathbb{R},dg_{\mu}\right)  }\left(
f\right)  \leq C_{3}E_{\mu,\lambda}\left(  B_{\mu}f\right)  +C_{4}\left\|
f\right\|  _{L^{2}\left(  \mathbb{R},dg_{\mu}\right)  }^{2}.\tag{1.4}%
\end{equation}

As is expected, by setting $\mu=0$ in (1.4) we can recover the inequality (1.2).

Finally, in Section 7 we present some conclusions and indicate some questions
that we have left unanswered in this work.

The first author has described in [Pi] a formalism which allows this theory to
be developed to the context of $\mathbb{R}^{n}$ and $\mathbb{C}^{n}$ in place
of $\mathbb{R}$ and $\mathbb{C}$. (See also [B-\O ].) We have not presented
this here, since the ideas are the same as in the case $n=1$ which we
consider. \bigskip

\section{Preliminaries}

\bigskip In this section we give the definitions and the notation (as well as
some preliminary results) that we will use throughout the work. First, we take
$\mu\geq0$ and $\lambda>0$ to be fixed parameters. The (Coxeter) group
$\mathbb{Z}_{2}$ is the multiplicative group $\left\{  -1,1\right\}  $, and
$\log$ is the natural logarithm (base $e$). We use the convention $0\log0=0$
(which makes the function $\phi:\left[  0,\infty\right)  \rightarrow\mathbb{R}
$, $\phi\left(  x\right)  =x\log x$ continuous). We also use the convention
that $C$ denotes a constant (a quantity that does not depend on the variables
of interest in the context), which may change its value every time it appears.
We will use when necessary (without further comment) the elementary inequality
$\left(  \alpha+\beta\right)  ^{r}\leq C_{r}\left(  \alpha^{r}+\beta
^{r}\right)  $, valid for all $r>0$ and $\alpha,\beta\geq0$. For two positive
functions $f$ and $g$ such that $\lim_{x\rightarrow a}\frac{f\left(  x\right)
}{g\left(  x\right)  }=1$, we use the notation $f\left(  x\right)  \cong
g\left(  x\right)  $ as $x\rightarrow a$. For a given $p\in\left[
1,+\infty\right]  $ we will denote by $p^{\prime}\in\left[  1,+\infty\right]
$ the Lebesgue dual index of $p$. We denote by $\mathcal{H}\left(
\mathbb{C}\right)  $ the space of holomorphic functions
$f:\mathbb{C\rightarrow C}$ with the topology of uniform convergence in
compact sets.

We begin by defining the $\mu$-deformed factorial function $\gamma_{\mu}$ and
$\mu$-deformed exponential function $\mathbf{e}_{\mu}$. Let $\mathbb{N}$
denote the set of positive integers.

\bigskip\textbf{Definition 2.1 }\textit{The $\mu$-deformed factorial function
}$\gamma_{\mu}:\mathbb{N}\cup\left\{  0\right\}  \rightarrow\mathbb{R}
$\textit{\ is defined by }$\gamma_{\mu}\left(  0\right)  =1$\textit{\ and }
\[
\gamma_{\mu}\left(  n\right)  :=\left(  n+2\mu\theta\left(  n\right)  \right)
\gamma_{\mu}\left(  n-1\right)  \mathit{, }%
\]
\textit{where }$n\in\mathbb{N}$\textit{ and }$\theta:\mathbb{N}\rightarrow
\left\{  0,1\right\}  $\textit{\ is the characteristic function of the odd
positive integers. The $\mu$-deformed } \textit{exponential function
}$\mathbf{e}_{\mu}:\mathbb{C}\rightarrow\mathbb{C}$\textit{, is defined by the
power series }
\[
\mathbf{e}_{\mu}\left(  z\right)  :=\sum_{n=0}^{\infty}\frac{z^{n}}%
{\gamma_{\mu}\left(  n\right)  }.
\]

\bigskip

We note that $\gamma_{0}\left(  n\right)  =n!$ and so $\mathbf{e}_{0}\left(
z\right)  =\exp\left(  z\right)  $. It is clear that the power series in the
definition of $\mathbf{e}_{\mu}\left(  z\right)  $ is absolutely convergent
for all $z\in\mathbb{C}$. So the $\mu$-deformed exponential $\mathbf{e}_{\mu}
$ is an entire function. Also note that $\gamma_{\mu}\left(  n\right)  \geq
n!$ (since we are assuming that $\mu\geq0$), and thus we have the inequality
$\mathbf{e}_{\mu}\left(  x\right)  \leq\exp\left(  x\right)  $ for all real
non-negative $x$.

In [Ros1] (Lemma 2.3) it is shown that for $\mu>0$ and $z\in\mathbb{C}$ one
has the following integral representation of the $\mu$-deformed exponential
function
\begin{equation}
\mathbf{e}_{\mu}(z)=\int_{-1}^{1}\exp(tz)d\sigma_{\mu}\left(  t\right)
,\tag{2.1}%
\end{equation}
where $d\sigma_{\mu} $ is the probability measure on $\left[  -1,1\right]  $
given by
\[
d\sigma_{\mu}\left(  t\right)  :=\frac{1}{B\left(  \frac{1}{2},\mu\right)
}\left(  1-t\right)  ^{\mu-1}\left(  1+t\right)  ^{\mu}dt
\]
and where $B$ is the beta function (see [Leb], p. 13). Note that $B(\frac
{1}{2} , \mu) > 0$ for $\mu> 0$. From this representation one gets easily the
fact that $\mathbf{e}_{\mu}\left(  x\right)  >0$ for all $x\in\mathbb{R}$.

\bigskip\textbf{Lemma 2.1 \ }\textit{For all }$\mu\geq0$\textit{\ and }%
$q\geq1$\textit{\ the following inequality holds for all $z \in\mathbb{C}$:}
\begin{equation}
\left\vert \mathbf{e}_{\mu}(z)\right\vert ^{q}\leq\mathbf{e}_{\mu}\left(
q\operatorname{Re}z\right)  .\tag{2.2}%
\end{equation}

\bigskip

\textbf{Proof: }Observe that if $\mu=0$ the inequality reduces to a trivial
equality for all $q\in\mathbb{R}$. If $q=1$ and $\mu>0$, one has, by using the
integral representation (2.1) of $\mathbf{e}_{\mu}\left(  z\right)  $, that
\[
\left\vert \mathbf{e}_{\mu}(z)\right\vert \leq\int_{-1}^{1}\left\vert
\exp(tz)\right\vert d\sigma_{\mu}\left(  t\right)  =\int_{-1}^{1}%
\exp(t\operatorname{Re}z)d\sigma_{\mu}\left(  t\right)  =\mathbf{e}_{\mu
}\left(  \operatorname{Re}z\right)  ,
\]
which proves the validity of the inequality for all $\mu>0$ and $q=1$. Thus,
it remains to prove the inequality in the case $\mu>0$ and $q>1$. Again by
using the integral representation (2.1) of $\mathbf{e}_{\mu}\left(  z\right)
$, H\"{o}lder's inequality, and the fact that $d\sigma_{\mu} $ is a
probability measure in $\left[  -1,1\right]  $, we have that
\begin{align*}
\left\vert \mathbf{e}_{\mu}(z)\right\vert  &  \leq\left(  \int_{-1}%
^{1}\left\vert \exp(tz)\right\vert ^{q}d\sigma_{\mu}\left(  t\right)  \right)
^{\frac{1}{q}}\left(  \int_{-1}^{1}d\sigma_{\mu}\left(  t\right)  \right)
^{\frac{1}{q^{\prime}}}\\
&  =\left(  \int_{-1}^{1}\exp(qt\operatorname{Re}z)d\sigma_{\mu}\left(
t\right)  \right)  ^{\frac{1}{q}}\\
&  =\left(  \mathbf{e}_{\mu}(q\operatorname{Re}z)\right)  ^{\frac{1}{q}},
\end{align*}
which proves the inequality in this case.

\hspace{0in}\hfill\textbf{Q.E.D.}

\bigskip

The following definition is due to Angulo and the second author (see [A-S.2]).

\bigskip

\textbf{Definition 2.2 }\textit{Let }$\lambda>0$\textit{. We define the
measure }$d\nu_{\mu,\lambda}$\textit{\ on the space }$\mathbb{C}%
\times\mathbb{Z}_{2}$ \textit{by}
\begin{equation}
d\nu_{\mu,\lambda}\left(  z,1\right)  :=\lambda\frac{2^{\frac{1}{2}-\mu}}%
{\pi\Gamma\left(  \mu+\frac{1}{2}\right)  }K_{\mu-\frac{1}{2}}\left(
\lambda\left\vert z\right\vert ^{2}\right)  \left\vert \lambda^{\frac{1}{2}%
}z\right\vert ^{2\mu+1}dxdy,\tag{2.3}%
\end{equation}
\begin{equation}
d\nu_{\mu,\lambda}\left(  z,-1\right)  :=\lambda\frac{2^{\frac{1}{2}-\mu}}%
{\pi\Gamma\left(  \mu+\frac{1}{2}\right)  }K_{\mu+\frac{1}{2}}\left(
\lambda\left\vert z\right\vert ^{2}\right)  \left\vert \lambda^{\frac{1}{2}%
}z\right\vert ^{2\mu+1}dxdy,\tag{2.4}%
\end{equation}
\textit{where }$\Gamma$\textit{\ is the Euler gamma function, }$K_{\alpha}%
$\textit{\ is the Macdonald function of order }$\alpha$\textit{\ (both defined
in }[Leb]\textit{), and }$dxdy$\textit{\ is Lebesgue measure on }$\mathbb{C}%
$\textit{.}

\bigskip

By convention, in the case $\lambda=1$ we will omit the parameter $\lambda$ in
the notation of the measure.

The Macdonald function $K_{\alpha}$ is the modified Bessel function of the
third kind (with purely imaginary argument, as described in [Wat], p. 78),
which is known to be a holomorphic function on $\mathbb{C}\setminus\left(
-\infty,0\right]  $ and is entire with respect to the parameter $\alpha$.
Nevertheless, our interest will be only in the values and behavior of this
function for $x\in\mathbb{R}^{+}$ and $\alpha\in\mathbb{R}$. For
$z\in\mathbb{C}$, $\left|  \arg z\right|  <\pi$ and $\alpha\notin\mathbb{Z}$,
the Macdonald function can be defined as
\[
K_{\alpha}\left(  z\right)  =\frac{\pi}{2}\frac{I_{-\alpha}\left(  z\right)
-I_{\alpha}\left(  z\right)  }{\sin\left(  \alpha\pi\right)  }%
\]
(see [Leb], p. 108), where $I_{\alpha}\left(  z\right)  $ is the modified
Bessel function of the first kind. For $\alpha\in\mathbb{Z}$, we define
$K_{\alpha}\left(  z\right)  =\lim_{\beta\rightarrow\alpha}K_{\beta}\left(
z\right)  $. This expression shows that $K_{a}\left(  z\right)  $ is an even
function of the parameter $\alpha$. In particular, since $I_{\frac{1}{2}%
}\left(  z\right)  =\left(  \frac{2}{\pi z}\right)  ^{\frac{1}{2}}\sinh z$ and
$I_{-\frac{1}{2}}\left(  z\right)  =\left(  \frac{2}{\pi z}\right)  ^{\frac
{1}{2}}\cosh z$ (see [Leb], p. 112), we have that%

\begin{equation}
K_{\pm\frac{1}{2}}\left(  z\right)  =\left(  \frac{\pi}{2z}\right)  ^{\frac
{1}{2}}\exp\left(  -z\right)  \text{,}\tag{2.5}%
\end{equation}
which shows that for $\mu=0$ the measures defined on $\mathbb{C}$ by (2.3) and
(2.4) are the same Gaussian measure:
\[
d\nu_{0,\lambda}\left(  z,1\right)  =d\nu_{0,\lambda}\left(  z,-1\right)
=\frac{\lambda}{\pi}\exp\left(  -\lambda\left\vert z\right\vert ^{2}\right)
dxdy.
\]

As is noted in [A-S.2], the last expression, when compared with the Gaussian
measure
\[
d\nu_{\text{Gauss,}\hbar}\left(  z\right)  :=\frac{1}{\pi\hbar}\exp\left(
-\frac{\left\vert z\right\vert ^{2}}{\hbar}\right)  dxdy,
\]
this being the measure of the Segal-Bargmann space, allows us to identify
$\lambda$ with $\hbar^{-1}$, where $\hbar>0$ is Planck's constant. (When
$\hbar^{-1}=\lambda=1$ we write this measure simply as $d\nu_{\text{Gauss}}
$.) We consider Planck's constant as a positive parameter. See [Hall], where
$\hbar$ is also identified with a ``time'' parameter denoted by $t$.

By using the formula
\[
\int_{0}^{\infty}K_{\alpha}\left(  s\right)  s^{\beta-1}ds=2^{\beta-2}%
\Gamma\left(  \frac{\beta-\alpha}{2}\right)  \Gamma\left(  \frac{\beta+\alpha
}{2}\right)  ,
\]
which holds if $\operatorname{Re}\beta>\left\vert \operatorname{Re}%
\alpha\right\vert $ (see [Wat], p. 388), we see that
\begin{align*}
\int_{\mathbb{C}}d\nu_{\mu,\lambda}\left(  z,1\right)   &  =\lambda
\frac{2^{\frac{1}{2}-\mu}}{\pi\Gamma\left(  \mu+\frac{1}{2}\right)  }%
\int_{\mathbb{C}}K_{\mu-\frac{1}{2}}\left(  \lambda\left\vert z\right\vert
^{2}\right)  \left\vert \lambda^{\frac{1}{2}}z\right\vert ^{2\mu+1}dxdy\\
&  =\frac{2^{\frac{1}{2}-\mu}}{\Gamma\left(  \mu+\frac{1}{2}\right)  }\int
_{0}^{\infty}K_{\mu-\frac{1}{2}}\left(  s\right)  s^{\mu+\frac{1}{2}}ds\\
&  =1,
\end{align*}
(where $s=\lambda r^{2}$, $r=\left\vert z\right\vert $), and
\begin{align*}
\int_{\mathbb{C}}d\nu_{\mu,\lambda}\left(  z,-1\right)   &  =\lambda
\frac{2^{\frac{1}{2}-\mu}}{\pi\Gamma\left(  \mu+\frac{1}{2}\right)  }%
\int_{\mathbb{C}}K_{\mu+\frac{1}{2}}\left(  \lambda\left\vert z\right\vert
^{2}\right)  \left\vert \lambda^{\frac{1}{2}}z\right\vert ^{2\mu+1}dxdy\\
&  =\frac{2^{\frac{1}{2}-\mu}}{\Gamma\left(  \mu+\frac{1}{2}\right)  }\int
_{0}^{\infty}K_{\mu+\frac{1}{2}}\left(  s\right)  s^{\mu+\frac{1}{2}}ds\\
&  =\frac{\pi^{\frac{1}{2}}\Gamma\left(  \mu+1\right)  }{\Gamma\left(
\mu+\frac{1}{2}\right)  }.
\end{align*}

That is, the measures $d\nu_{\mu,\lambda}\left(  z,1\right)  $ and $d\nu
_{\mu,\lambda}\left(  z,-1\right)  $ on $\mathbb{C}$ are finite, and moreover
the former is a probability measure. Another way of seeing this is given in [A-S.2].

The integral representation
\[
K_{\alpha}\left(  z\right)  =\int_{0}^{\infty}\exp\left(  -z\cosh u\right)
\cosh\left(  \alpha u\right)  du\qquad\operatorname{Re}z>0
\]
(see [Leb], p. 119) gives us at once two important properties of the Macdonald
function. The first is that $K_{\alpha}\left(  x\right)  >0$ for all
$x\in\mathbb{R}^{+}$, and the second is that $K_{\alpha}$ is a monotone
decreasing function for $x\in\mathbb{R}^{+}$.

We will use the following facts about the asymptotic behavior of the Macdonald
function (see [Leb], pp. 110,136):
\begin{equation}
K_{\alpha}\left(  x\right)  \cong\frac{2^{\left\vert \alpha\right\vert
-1}\Gamma\left(  \left\vert \alpha\right\vert \right)  }{x^{\left\vert
\alpha\right\vert }}\qquad\text{as}\qquad x\rightarrow0^{+}\qquad
\text{if}\qquad\alpha\neq0.\tag{2.6}%
\end{equation}%
\begin{equation}
K_{0}\left(  x\right)  \cong\log\frac{2}{x}\qquad\text{as}\qquad
x\rightarrow0^{+}.\tag{2.7}%
\end{equation}%
\begin{equation}
K_{\alpha}\left(  x\right)  \cong\left(  \frac{\pi}{2x}\right)  ^{\frac{1}{2}%
}\exp\left(  -x\right)  \qquad\text{as}\qquad x\rightarrow+\infty
\qquad\text{for all}\qquad\alpha\in\mathbb{R}.\tag{2.8}%
\end{equation}

We will be dealing with the complex Banach spaces $L^{p}\left(  \Omega
,d\nu\right)  $ where $\left(  \Omega,d\nu\right)  $ is a measure space and
$1\leq p\leq\infty$. In fact, the measure spaces $\left(  \Omega,d\nu\right)
$ involved in this work will always be finite. We will denote the norm of a
vector $f\in L^{p}\left(  \Omega,d\nu\right)  $ by $\left\Vert f\right\Vert
_{L^{p}\left(  \Omega,d\nu\right)  }$. If $\left(  \Omega_{i},d\nu_{i}\right)
$, $i=1,2$ are measure spaces and $p,q\geq1$, the norm of an operator defined
in some dense subspace $D$ of $L^{p}\left(  \Omega_{1},d\nu_{1}\right)  $ with
image in $L^{q}\left(  \Omega_{2},d\nu_{2}\right)  $ is defined by
\[
\left\Vert T\right\Vert _{p\rightarrow q}:=\sup\left\{  \left\Vert
Tf\right\Vert _{L^{q}\left(  \Omega_{2},d\nu_{2}\right)  }:f\in D\text{,
}\left\Vert f\right\Vert _{L^{p}\left(  \Omega_{1},d\nu_{1}\right)
}=1\right\}  .
\]

This is the \textit{operator norm} of $T$. Although the corresponding measure
spaces $\left(  \Omega_{i},d\nu_{i}\right)  $, $i=1,2$, do not appear in the
notation $\left\|  T\right\|  _{p\rightarrow q}$, these spaces will be clear
from context.

The most important operators we will deal with in this work are operators $T$
from some dense domain $D$ of a space $L^{p}\left(  X,d\rho\right)  $ into
some space $L^{q}\left(  Y,d\sigma\right)  $ (where $\left(  X,d\rho\right)  $
and $\left(  Y,d\sigma\right)  $ are finite measure spaces), which are
integral kernel operators of the form
\[
\left(  Tf\right)  \left(  y\right)  =\int_{X}\widetilde{T}\left(  x,y\right)
f\left(  x\right)  d\rho\left(  x\right)  ,
\]
where $\widetilde{T}:X\times Y\rightarrow\mathbb{C}$ is a measurable function,
called the \textit{kernel of the operator} $T$ and usually denoted by the same
letter $T$. We define the \textit{Hille-Tamarkin norm} of the kernel $T$,
denoted by $|||T|||_{p,q}$ (unfortunately with the same ambiguity as that of
the operator norm), by
\begin{equation}
|||T|||_{p,q}:=\left\Vert T_{p}\right\Vert _{L^{q}\left(  Y,d\sigma\right)
}\text{,}\tag{2.9}%
\end{equation}
where $T_{p}\left(  y\right)  =\left\Vert T\left(  \cdot,y\right)  \right\Vert
_{L^{p^{\prime}}\left(  X,d\rho\right)  }$, $y\in Y$. If $1<p\leq\infty$ and
$1\leq q<\infty$, we explicitly have
\[
|||T|||_{p,q}=\left\{  \int_{Y}\left(  \int_{X}\left\vert T\left(  x,y\right)
\right\vert ^{p^{\prime}}d\rho\left(  x\right)  \right)  ^{\frac{q}{p^{\prime
}}}d\sigma\left(  y\right)  \right\}  ^{\frac{1}{q}}.
\]

(Note that $||| T |||_{2,2}$ is the Hilbert-Schmidt norm of $T$.)

Given a pair of Lebesgue indices $\left(  p,q\right)  \in\left[
1,\infty\right]  \times\left[  1,\infty\right]  $, we say that the integral
kernel operator $T$ (as described above) is a \textit{Hille-Tamarkin operator}
with respect to the pair $\left(  p,q\right)  $ if the Hille-Tamarkin norm
(2.9) is finite. It can be proved that the set of Hille-Tamarkin operators
with respect to $\left(  p,q\right)  $ is a complex vector space, that (2.9)
defines a norm on it, and that this normed space is in fact a Banach space
(see Theorem 11.5 of [J]).

We will use also the following two results (see [J], Theorems 11.5 and 11.6).

\bigskip\textbf{Proposition 2.1}\textit{\ \ }$\left\|  T\right\|
_{p\rightarrow q}\leq||| T |||_{p,q}$.

\bigskip

This proposition tells us that the Hille-Tamarkin operators with respect a
given pair $\left(  p,q\right)  $ are bounded from $L^{p}\left(
X,d\rho\right)  $ to $L^{q}\left(  Y,d\sigma\right)  $.

\bigskip

\textbf{Proposition 2.2}\textit{\ \ If }$||| T |||_{p,q}<\infty$\textit{\ and
}$1\leq q<\infty$\textit{\ and }$1<p\leq\infty$\textit{, then }$T$\textit{\ is
a compact operator from }$L^{p}\left(  X,d\rho\right)  $ \textit{to }%
$L^{q}\left(  Y,d\sigma\right)  $.

\bigskip

We will work with the Banach space $L^{p}\left(  \mathbb{C}\times
\mathbb{Z}_{2},d\nu_{\mu,\lambda}\right)  $, where $p\geq1$.

Let us consider the space
\[
\mathfrak{B}_{p,\mu,\lambda}\!\!=\!\!\left\{  f:\mathbb{C}\rightarrow
\mathbb{C}\mid f_{e}\in L^{p}( \mathbb{C},\left.  d\nu_{\mu,\lambda}\right|
_{\mathbb{C}\times\left\{  1\right\}  }) \text{ and }f_{o}\in L^{p}(
\mathbb{C},\left.  d\nu_{\mu,\lambda}\right|  _{\mathbb{C}\times\left\{
-1\right\}  }) \right\}  ,
\]
where $f=f_{e}+f_{o}$ is the decomposition of $f$ in its even and odd parts.
Here and subsequently we identify these two restrictions of $d\nu_{\mu
,\lambda}$ as measures on $\mathbb{C}$, using $\mathbb{C\cong C\times}\left\{
1\right\}  \mathbb{\cong C\times}\left\{  -1\right\}  $. Moreover, we will use
without further comment the notation $f_{e}$ ($f_{o}$) for the even part (the
odd part, respectively) of a function $f$.

For $p\geq1$ and $f\in\mathfrak{B}_{p,\mu,\lambda}$ we define
\[
\left\Vert f\right\Vert _{\mathfrak{B}_{p,\mu,\lambda}}^{p}:=\left\Vert
f_{e}\right\Vert _{L^{p}\left(  \mathbb{C},\left.  d\nu_{\mu,\lambda
}\right\vert _{\mathbb{C}\times\left\{  1\right\}  }\right)  }^{p}+\left\Vert
f_{o}\right\Vert _{L^{p}\left(  \mathbb{C},\left.  d\nu_{\mu,\lambda
}\right\vert _{\mathbb{C}\times\left\{  -1\right\}  }\right)  }^{p}.
\]

The linear map $\Phi:\mathfrak{B}_{p,\mu,\lambda}\rightarrow L^{p}\left(
\mathbb{C}\times\mathbb{Z}_{2},d\nu_{\mu,\lambda}\right)  $ defined as
$\left(  \Phi f\right)  \left(  z,1\right)  =f_{e}\left(  z\right)  $ and
$\left(  \Phi f\right)  \left(  z,-1\right)  =f_{o}\left(  z\right)  $ is
injective and has the property that
\begin{equation}
\left\Vert f\right\Vert _{\mathfrak{B}_{p,\mu,\lambda}}=\left\Vert \Phi
f\right\Vert _{L^{p}\left(  \mathbb{C}\times\mathbb{Z}_{2},d\nu_{\mu,\lambda
}\right)  }\tag{2.10}%
\end{equation}
for all $f\in\mathfrak{B}_{p,\mu,\lambda}$. Therefore $\left\Vert
\cdot\right\Vert _{\mathfrak{B}_{p,\mu,\lambda}}$ is a norm on $\mathfrak{B}%
_{p,\mu,\lambda}$. It is not hard to show that the range of $\Phi$ is a closed
subspace of $L^{p}\left(  \mathbb{C}\times\mathbb{Z}_{2},d\nu_{\mu,\lambda
}\right)  $ for $p\geq1$. (The proof is similar to one found in [Hall].)
Therefore $\mathfrak{B}_{p,\mu,\lambda}$ is a Banach space, since we have
identified it with a closed subspace of the Banach space $L^{p}\left(
\mathbb{C}\times\mathbb{Z}_{2},d\nu_{\mu,\lambda}\right)  $. For a function
$f\in\mathfrak{B}_{p,\mu,\lambda}$ we will sometimes write its norm as
$\left\Vert f\right\Vert _{L^{p}\left(  \mathbb{C}\times\mathbb{Z}_{2}%
,d\nu_{\mu,\lambda}\right)  }$, meaning that we are using (2.10) and
identifying $f$ with $\Phi f$.

We will use the notations $d\nu_{e,\mu,\lambda}$ and $d\nu_{o,\mu,\lambda}$
for the restrictions $\left.  d\nu_{\mu,\lambda}\right|  _{\mathbb{C}%
\times\left\{  1\right\}  }$ and $\left.  d\nu_{\mu,\lambda}\right|
_{\mathbb{C}\times\left\{  -1\right\}  }$, respectively. So for $f\in
\mathfrak{B}_{p,\mu,\lambda}$ we have
\begin{align*}
\left\|  f\right\|  _{\mathfrak{B}_{p,\mu,\lambda}}^{p}  &  =\left\|
f_{e}\right\|  _{L^{p}\left(  \mathbb{C},d\nu_{e,\mu,\lambda}\right)  }%
^{p}+\left\|  f_{o}\right\|  _{L^{p}\left(  \mathbb{C},d\nu_{o,\mu,\lambda
}\right)  }^{p}\\
&  =\left\|  f_{e}\right\|  _{\mathfrak{B}_{p,\mu,\lambda}}^{p}+\left\|
f_{o}\right\|  _{\mathfrak{B}_{p,\mu,\lambda}}^{p}.
\end{align*}

Observe that this says that $\mathfrak{B}_{p,\mu,\lambda}=\mathfrak{B}%
_{e,p,\mu,\lambda}\oplus\mathfrak{B}_{o,p,\mu,\lambda}$, where
\[
\mathfrak{B}_{e,p,\mu,\lambda}=\left\{  f\in\mathfrak{B}_{p,\mu,\lambda}\text{
}\mid f=\text{ }f_{e}\right\}
\]
and
\[
\mathfrak{B}_{o,p,\mu,\lambda}=\left\{  f\in\mathfrak{B}_{p,\mu,\lambda}\text{
}\mid f=\text{ }f_{o}\right\}
\]
are Banach subspaces of $\mathfrak{B}_{p,\mu,\lambda}$.

Let us consider the dilation operator $T_{\lambda}\left(  f\right)  \left(
z\right)  =f\left(  \lambda^{\frac{1}{2}}z\right)  $. Let us see that
$T_{\lambda}$ is an isometry from $\mathfrak{B}_{p,\mu}$ onto $\mathfrak{B}%
_{p,\mu,\lambda}$. Observe that
\begin{align}
&  \int_{\mathbb{C}}\left\vert f\left(  z\right)  \right\vert ^{p}K_{\mu
-\frac{1}{2}}\left(  \left\vert z\right\vert ^{2}\right)  \left\vert
z\right\vert ^{2\mu+1}dxdy\nonumber\\
&  =\ \ {}\int_{\mathbb{C}}\left\vert f\left(  \lambda^{\frac{1}{2}}z\right)
\right\vert ^{p}\lambda K_{\mu-\frac{1}{2}}\left(  \lambda\left\vert
z\right\vert ^{2}\right)  \left\vert \lambda^{\frac{1}{2}}z\right\vert
^{2\mu+1}dxdy\nonumber
\end{align}
by a change of variables. This shows that $T_{\lambda}f\in\mathfrak{B}%
_{e,p,\mu,\lambda}$ if and only if $f\in\mathfrak{B}_{e,p,\mu}$, and moreover,
that $\left\Vert f\right\Vert _{\mathfrak{B}_{p,\mu}}=\left\Vert T_{\lambda
}f\right\Vert _{\mathfrak{B}_{p,\mu,\lambda}}$. Similarly, $T_{\lambda}%
f\in\mathfrak{B}_{o,p,\mu,\lambda}$ if and only if $f\in\mathfrak{B}_{o,p,\mu
}$, and $\left\Vert f\right\Vert _{\mathfrak{B}_{p,\mu}}=\left\Vert
T_{\lambda}f\right\Vert _{\mathfrak{B}_{p,\mu,\lambda}}$. Since clearly
$\left(  T_{\lambda}f\right)  _{e}=T_{\lambda}\left(  f_{e}\right)  $ and
$\left(  T_{\lambda}f\right)  _{o}=T_{\lambda}\left(  f_{o}\right)  $, we have
that
\begin{align*}
\left\Vert f\right\Vert _{\mathfrak{B}_{p,\mu}}^{p} &  =\left\Vert
f_{e}\right\Vert _{\mathfrak{B}_{p,\mu}}^{p}+\left\Vert f_{o}\right\Vert
_{\mathfrak{B}_{p,\mu}}^{p}\\
&  =\left\Vert T_{\lambda}\left(  f_{e}\right)  \right\Vert _{\mathfrak{B}%
_{p,\mu,\lambda}}^{p}+\left\Vert T_{\lambda}\left(  f_{o}\right)  \right\Vert
_{\mathfrak{B}_{p,\mu,\lambda}}^{p}\\
&  =\left\Vert T_{\lambda}f\right\Vert _{\mathfrak{B}_{p,\mu,\lambda}}^{p},
\end{align*}
which proves our claim. In particular, when $p=2$, the dilation operator
$T_{\lambda}$ is unitary.

\bigskip

\textbf{Definition 2.3 \ }\textit{Let }$\left(  \Omega,d\nu\right)
$\textit{\ be a finite measure space, that is, $0<\nu(\Omega)<\infty$. For
}$f\in L^{2}\left(  \Omega,d\nu\right)  $\textit{, the entropy }%
$S_{L^{2}\left(  \Omega,d\nu\right)  }\left(  f\right)  $\textit{\ is defined
by }
\begin{equation}
S_{L^{2}\left(  \Omega,d\nu\right)  }\left(  f\right)  :=\int_{\Omega
}\left\vert f\left(  \omega\right)  \right\vert ^{2}\log\left\vert f\left(
\omega\right)  \right\vert ^{2}d\nu\left(  \omega\right)  -\left\Vert
f\right\Vert _{L^{2}\left(  \Omega,d\nu\right)  }^{2}\log\left\Vert
f\right\Vert _{L^{2}\left(  \Omega,d\nu\right)  }^{2}.\tag{2.11}%
\end{equation}

\bigskip This definition was introduced by Shannon [Sha] in his Theory of
Communication. Note that, since $\left(  \Omega,d\nu\right)  $ is a finite
measure space, the entropy $S_{L^{2}\left(  \Omega,d\nu\right)  }\left(
f\right)  $ makes sense for all $f\in L^{2}\left(  \Omega,d\nu\right)  $.
Moreover, by considering the convex function $\phi:\left[  0,\infty\right)
\rightarrow\mathbb{R}$, $\phi\left(  x\right)  =x\log x$, and the probability
measure space $\left(  \Omega,d\nu^{\prime}\right)  $, where $d\nu^{\prime
}=W^{-1}d\nu$, $W=\nu(\Omega)$, we have by Jensen's inequality (see [L-L], p.
38) that
\[
\left(  \int_{\Omega}\left\vert f\left(  \omega\right)  \right\vert ^{2}%
d\nu\left(  \omega\right)  \right)  \!\log\!\left(  \frac{1}{W}\int_{\Omega
}\left\vert f\left(  \omega\right)  \right\vert ^{2}d\nu\left(  \omega\right)
\right)  \!\leq\!\int_{\Omega}\left\vert f\left(  \omega\right)  \right\vert
^{2}\log\left\vert f\left(  \omega\right)  \right\vert ^{2}d\nu\left(
\omega\right)
\]
or
\[
\left(  -\log W\right)  \left\Vert f\right\Vert _{L^{2}\left(  \Omega
,d\nu\right)  }^{2}\leq S_{L^{2}\left(  \Omega,d\nu\right)  }\left(  f\right)
,
\]
which shows that $S_{L^{2}\left(  \Omega,d\nu\right)  }\left(  f\right)
\neq-\infty$, though $S_{L^{2}\left(  \Omega,d\nu\right)  }\left(  f\right)
=+\infty$ can happen. Also observe that $S_{L^{2}\left(  \Omega,d\nu^{\prime
}\right)  }\left(  f\right)  \geq0$, though $S_{L^{2}\left(  \Omega
,d\nu\right)  }\left(  f\right)  $ can be negative. Finally, note that
$S_{L^{2}\left(  \Omega,d\nu\right)  }\left(  f\right)  $ is homogeneous of
degree $2$.

\bigskip

\section{The $\lambda$-weighted $\mu$-deformed Segal-Bargmann \newline space
and its transform}

\bigskip

We begin by defining the Segal-Bargmann space of interest for us in this work.

\bigskip

\textbf{Definition 3.1 }\textit{Let }$1\leq q<\infty$\textit{. The } $\lambda
$-\textit{weighted} $\mu$-\textit{deformed Segal-Bargmann space, denoted by}
$\mathcal{B}_{\mu,\lambda}^{q}$\textit{, is defined as}
\[
\mathcal{B}_{\mu,\lambda}^{q}:=\mathcal{H}\left(  \mathbb{C}\right)
\cap\mathfrak{B}_{q,\mu,\lambda}.
\]

\bigskip

Although this definition makes sense for $0<q<\infty$, we will only be
interested in the case $1\leq q<\infty$, since in this case the space
$\mathcal{B}_{\mu,\lambda}^{q}$ (the holomorphic subspace of the Banach space
$\mathfrak{B}_{q,\mu,\lambda}$) is a Banach space with the norm of
$\mathfrak{B}_{q,\mu,\lambda}$.

If we decompose the space $\mathcal{H}\left(  \mathbb{C}\right)  $ of
holomorphic functions $f:\mathbb{C\rightarrow C}$, as $\mathcal{H}\left(
\mathbb{C}\right)  =\mathcal{H}_{e}\left(  \mathbb{C}\right)  \oplus
\mathcal{H}_{o}\left(  \mathbb{C}\right)  $, where
\begin{align*}
\mathcal{H}_{e}\left(  \mathbb{C}\right)  \!\!  &  :=\left\{  f\in
\mathcal{H}\left(  \mathbb{C}\right)  :f=f_{e}\right\} \\
\text{and} \ \ \mathcal{H}_{o}\left(  \mathbb{C}\right)  \!\!  &  :=\left\{
f\in\mathcal{H}\left(  \mathbb{C}\right)  :f=f_{o}\right\}
\end{align*}
are the subspaces of the even and odd functions of $\mathcal{H}\left(
\mathbb{C}\right)  $, respectively, then by writing $\mathcal{H}\left(
\mathbb{C}\right)  \ni f=f_{e}+f_{o}$, the space $\mathcal{B}_{\mu,\lambda
}^{q}$ is just the space of holomorphic functions $f:\mathbb{C\rightarrow C}$
such that the even part $f_{e}$ (the odd part $f_{o}$) of $f$ is $q$
integrable with respect to the measure $d\nu_{e,\mu,\lambda}$ (with respect to
the measure $d\nu_{o,\mu,\lambda}$, respectively). That is,
\[
\mathcal{B}_{\mu,\lambda}^{q}=\left\{  f\in\mathcal{H}\left(  \mathbb{C}%
\right)  :f_{e}\in L^{q}\left(  \mathbb{C},d\nu_{e,\mu,\lambda}\right)  \text{
and }f_{o}\in L^{q}\left(  \mathbb{C},d\nu_{o,\mu,\lambda}\right)  \right\}  .
\]

Yet another way to think of $\mathcal{B}_{\mu,\lambda}^{q}$ is as
\[
\mathcal{B}_{\mu,\lambda}^{q}=\mathcal{B}_{e,\mu,\lambda}^{q}\oplus
\mathcal{B}_{o,\mu,\lambda}^{q},
\]
where
\begin{align*}
\mathcal{B}_{e,\mu,\lambda}^{q}\!\! &  =\!\!\mathcal{H}\left(  \mathbb{C}%
\right)  \cap\mathfrak{B}_{e,q,\mu,\lambda}=\mathcal{H}_{e}\left(
\mathbb{C}\right)  \cap\mathfrak{B}_{q,\mu,\lambda}\\
\text{and}\ \ \ \ \mathcal{B}_{o,\mu,\lambda}^{q}\!\! &  =\!\!\mathcal{H}%
\left(  \mathbb{C}\right)  \cap\mathfrak{B}_{o,q,\mu,\lambda}=\mathcal{H}%
_{o}\left(  \mathbb{C}\right)  \cap\mathfrak{B}_{q,\mu,\lambda}%
\end{align*}
are the even and odd subspaces of $\mathcal{B}_{\mu,\lambda}^{q}$.

In the case $q=2$, the inner product of the Hilbert space $\mathcal{B}%
_{\mu,\lambda}^{2}$ (from which the norm on $\mathcal{B}_{\mu,\lambda}^{2}$
defined above comes) is
\[
\left\langle f,g\right\rangle _{\mathcal{B}_{\mu,\lambda}^{2}}=\left\langle
f_{e},g_{e}\right\rangle _{L^{2}\left(  \mathbb{C},d\nu_{e,\mu,\lambda
}\right)  }+\left\langle f_{o},g_{o}\right\rangle _{L^{2}\left(
\mathbb{C},d\nu_{o,\mu,\lambda}\right)  }.
\]

We then have that the even subspace $\mathcal{B}_{e,\mu,\lambda}^{2}$ of the
space $\mathcal{B}_{\mu,\lambda}^{2}$ is orthogonal to its odd subspace
$\mathcal{B}_{o,\mu,\lambda}^{q}$, and $\mathcal{B}_{\mu,\lambda}%
^{2}=\mathcal{B}_{e,\mu,\lambda}^{2}\oplus\mathcal{B}_{o,\mu,\lambda}^{2}$ as
Hilbert spaces. When $\mu=0$ and $\lambda=1$ we have the Segal-Bargmann space
$\mathcal{B}^{2}=\mathcal{H}\left(  \mathbb{C}\right)  \cap L^{2}\left(
\mathbb{C},d\nu_{\text{Gauss}}\right)  $ that appears in the \textquotedblleft
undeformed\textquotedblright\ theory (see [Hall]).

Observe that $T_{\lambda}:\mathcal{B}_{\mu}^{2}\rightarrow\mathcal{B}%
_{\mu,\lambda}^{2}$, $\left(  T_{\lambda}f\right)  \left(  z\right)  =f\left(
\lambda^{\frac{1}{2}}z\right)  $, is a unitary operator. This comes from the
fact that the dilation operator $T_{\lambda}:\mathfrak{B}_{2,\mu}%
\rightarrow\mathfrak{B}_{2,\mu,\lambda}$ is unitary (as we proved in the
previous section), and the fact that $T_{\lambda}f\in\mathcal{H}\left(
\mathbb{C}\right)  $ if and only if $f\in\mathcal{H}\left(  \mathbb{C}\right)
$.

The space $\mathcal{B}_{\mu}^{2}$ with $\mu>-\frac{1}{2}$ was studied by
Rosenblum ([Ros2]) and by Marron ([Marr]). It is known that $\left\{  \xi
_{n}^{\mu}\right\}  _{n=0}^{\infty}$, where $\xi_{n}^{\mu}\left(  z\right)
:=\left(  \gamma_{\mu}\left(  n\right)  \right)  ^{-\frac{1}{2}}z^{n}$,$\ $is
an orthonormal basis of $\mathcal{B}_{\mu}^{2}$ (see [Marr], p. 15, and
[A-S.1]). It follows that $\left\{  \chi_{n}^{\mu}\right\}  _{n=0}^{\infty}$,
where $\chi_{n}^{\mu}\left(  z\right)  :=\left(  \gamma_{\mu}\left(  n\right)
\right)  ^{-\frac{1}{2}}\lambda^{\frac{n}{2}}z^{n}$, is an orthonormal basis
of $\mathcal{B}_{\mu,\lambda}^{2}$, which is obtained by applying the dilation
operator $T_{\lambda}:\mathcal{B}_{\mu}^{2}\rightarrow\mathcal{B}_{\mu
,\lambda}^{2}$ to the elements of the basis $\left\{  \xi_{n}^{\mu}\right\}
_{n=0}^{\infty}$.

\bigskip

Rosenblum and Marron considered the \textit{$\mu$-deformed Bargmann transform}
$\widetilde{B}_{\mu}:L^{2}\left(  \mathbb{R},\left\vert t\right\vert ^{2\mu
}dt\right)  \rightarrow\mathcal{B}_{\mu}^{2}$ (which they called the
\textit{generalized Segal-Bargmann transform}). This can be defined by
$\widetilde{B}_{\mu}\left(  \phi_{n}^{\mu}\right)  =\xi_{n}^{\mu}$, where
$\left\{  \xi_{n}^{\mu}\right\}  _{n=0}^{\infty}$ is the orthonormal basis of
the $\mu$-deformed Segal-Bargmann space $\mathcal{B}_{\mu}^{2}$ mentioned
above, and $\left\{  \phi_{n}^{\mu}\right\}  _{n=0}^{\infty}$ is the
orthonormal basis of $L^{2}\left(  \mathbb{R},\left\vert t\right\vert ^{2\mu
}dt\right)  $ formed by the \textit{$\mu$-deformed Hermite functions }%
$\phi_{n}^{\mu}$ defined by
\[
\phi_{n}^{\mu}(t):=\left(  \frac{\gamma_{\mu}(n)}{\Gamma\left(  \mu+\frac
{1}{2}\right)  }\right)  ^{\frac{1}{2}}\frac{1}{2^{\frac{n}{2}}n!}\exp\left(
-\frac{t^{2}}{2}\right)  H_{n}^{\mu}(t),
\]
where $H_{n}^{\mu}\left(  t\right)  $ is the $n$-\textit{th $\mu$-deformed
Hermite polynomial} defined by the generating function
\[
\exp\left(  -z^{2}\right)  \mathbf{e}_{\mu}\left(  2tz\right)  =\sum
_{n=0}^{\infty}H_{n}^{\mu}\left(  t\right)  \frac{z^{n}}{n!}.
\]
(It is easy to check that $H_{n}^{\mu}\left(  t\right)  $ so defined is in
fact a polynomial of degree $n$ in $t$.) Clearly $\widetilde{B}_{\mu}$ is a
unitary map from the $\mu$-deformed quantum configuration space $L^{2}\left(
\mathbb{R},\left\vert t\right\vert ^{2\mu}dt\right)  $ onto the $\mu$-deformed
quantum phase space $\mathcal{B}_{\mu}^{2}$. We mention that the parameter
$\mu$ in the work of Rosenblum and Marron takes values in $\left(  -\frac
{1}{2},+\infty\right)  $, and not only in $\left[  0,+\infty\right)  $ as we
are considering in this work. As far as we know, the inequality of Lemma 2.1
is valid only for non-negative values of $\mu$. This lemma is used in the
proof of the main result of the next section (Theorem 4.1), and this result in
turn plays a fundamental role in the statement and proof of the theorems of
Sections 5 and 6.

An explicit formula for $\widetilde{B}_{\mu}$ is (see [Marr], p. 16)
\[
\left(  \widetilde{B}_{\mu}f\right)  \left(  z\right)  =\frac{1}{\left(
\Gamma\left(  \mu+\frac{1}{2}\right)  \right)  ^{\frac{1}{2}}}\exp\left(
-\frac{z^{2}}{2}\right)  \int_{\mathbb{R}}f\left(  t\right)  \mathbf{e}_{\mu
}\left(  2^{\frac{1}{2}}tz\right)  \exp\left(  -\frac{t^{2}}{2}\right)
\left\vert t\right\vert ^{2\mu}dt.
\]

The point of view we will adopt here (as in [Snt1]) is to replace the $\mu
$-deformed quantum configuration space $L^{2}\left(  \mathbb{R},\left\vert
t\right\vert ^{2\mu}dt\right)  $ by another unitarily equivalent space
$L^{2}\left(  \mathbb{R},dg_{\mu}\right)  $, known as the \textit{$\mu
$-deformed ground state representation}, where $dg_{\mu}(t):=\left(  \phi
_{0}^{\mu}(t)\right)  ^{2}\left\vert t\right\vert ^{2\mu}dt$ and $\phi
_{0}^{\mu}(t)=\left(  \Gamma\left(  \mu+\frac{1}{2}\right)  \right)
^{-\frac{1}{2}}\exp\left(  -\frac{t^{2}}{2}\right)  $ is the \textit{ground
state} (the first element of the orthonormal basis $\left\{  \phi_{n}^{\mu
}\right\}  _{n=0}^{\infty}$ of $L^{2}\left(  \mathbb{R},\left\vert
t\right\vert ^{2\mu}dt\right)  $ mentioned above). Notice that $dg_{\mu}$ is a
probability measure that generalizes the Gaussian probability measure
$dg\left(  t\right)  :=\pi^{-\frac{1}{2}}\exp\left(  -t^{2}\right)  dt$ that
appears in the case $\mu=0$ (see [Hall], p. 25). Explicitly $dg_{\mu}$ looks
like
\begin{equation}
dg_{\mu}(t)=\left(  \Gamma\left(  \mu+\frac{1}{2}\right)  \right)  ^{-1}%
\exp\left(  -t^{2}\right)  \left\vert t\right\vert ^{2\mu}dt.\tag{3.1}%
\end{equation}

Also, it is clear that $G:L^{2}\left(  \mathbb{R},\left\vert t\right\vert
^{2\mu}dt\right)  \rightarrow L^{2}\left(  \mathbb{R},dg_{\mu}\right)  $
defined as
\[
\left(  Gf\right)  \left(  t\right)  =\left(  \Gamma\left(  \mu+\frac{1}%
{2}\right)  \right)  ^{\frac{1}{2}}\exp\left(  \frac{t^{2}}{2}\right)
f\left(  t\right)  =\frac{f\left(  t\right)  }{\phi_{0}^{\mu}(t)}%
\]
is a unitary onto map, and then $B_{\mu}=\widetilde{B}_{\mu}\circ G^{-1}%
:L^{2}\left(  \mathbb{R},dg_{\mu}\right)  \rightarrow\mathcal{B}_{\mu}^{2}$ is
also a unitary map from the $\mu$-deformed ground state representation
$L^{2}\left(  \mathbb{R},dg_{\mu}\right)  $ onto the $\mu$-deformed
Segal-Bargmann space $\mathcal{B}_{\mu}^{2}$. It is easy to see, from the
explicit formula for $\widetilde{B}_{\mu}$ and (3.1), that an explicit formula
for $B_{\mu}$ is
\[
\left(  B_{\mu}f\right)  \left(  z\right)  =\exp\left(  -\frac{z^{2}}%
{2}\right)  \int_{\mathbb{R}}\mathbf{e}_{\mu}\left(  2^{\frac{1}{2}}tz\right)
f\left(  t\right)  dg_{\mu}\left(  t\right)  .
\]

We will call the transform $B_{\mu}:L^{2}\left(  \mathbb{R},dg_{\mu}\right)
\rightarrow\mathcal{B}_{\mu}^{2}$, defined by the formula above, the
\textit{$\mu$-deformed Segal-Bargmann transform }. Observe that if we set
$\mu=0$ this formula becomes
\[
\left(  B_{0}f\right)  (z)=\int_{\mathbb{R}}\exp\left(  -\frac{z^{2}}%
{2}+2^{\frac{1}{2}}tz\right)  f(t)dg\left(  t\right)  ,
\]
which is the \textquotedblleft usual\textquotedblright\ Segal-Bargmann
transform studied, for example, in [Hall], where it is shown that is a unitary
map from the quantum configuration space $L^{2}\left(  \mathbb{R},dg\right)  $
(the ground-state representation) onto the quantum phase space $\mathcal{B}%
^{2}=\mathcal{H}\left(  \mathbb{C}\right)  \cap L^{2}\left(  \mathbb{C}%
,d\nu_{\text{Gauss}}\right)  $ (the Segal-Bargmann space).

For example, let us consider the function $f_{n}\left(  t\right)  =t^{n}$
which lies in $L^{2}\left(  \mathbb{R},dg_{\mu}\right)  $ for any integer
$n\geq0$. The $\mu$-deformed Segal-Bargmann transform of $f_{n}$ is
\[
\left(  B_{\mu}f_{n}\right)  \left(  z\right)  =\left(  \Gamma\left(
\mu+\frac{1}{2}\right)  \right)  ^{-1}\exp\left(  -\frac{z^{2}}{2}\right)
\int_{\mathbb{R}}\mathbf{e}_{\mu}\left(  2^{\frac{1}{2}}tz\right)  t^{n}%
\exp\left(  -t^{2}\right)  \left\vert t\right\vert ^{2\mu}dt.
\]

To evaluate this we will use
\[
\int_{\mathbb{R}}\mathbf{e}_{\mu}\left(  -ixt\right)  t^{n}\exp\left(
-t^{2}\right)  \left\vert t\right\vert ^{2\mu}dt= \frac{\left(  {-i}\right)
^{n}\Gamma\left(  \mu+\frac{1}{2}\right)  \gamma_{\mu}\left(  n\right)  }%
{{2}^{n}n!}\exp\left(  -\frac{x^{2}}{4}\right)  H_{n}^{\mu}\left(  \frac{x}%
{2}\right)
\]
(see [Ros1], p. 378). Then we have that
\[
\left(  B_{\mu}f_{n}\right)  \left(  z\right)  =\frac{\gamma_{\mu}\left(
n\right)  }{n!}\left(  -\frac{i}{2}\right)  ^{n}H_{n}^{\mu}\left(
2^{-\frac{1}{2}}iz\right)  .
\]

For example, if $n=0$ we have $H_{0}^{\mu}\left(  t\right)  =1$ and then
$\left(  B_{\mu}f_{0}\right)  \left(  z\right)  =1$. If $n=1$ we have
$H_{1}^{\mu}\left(  t\right)  =\frac{2}{1+2\mu}t$ and then $\left(  B_{\mu
}f_{1}\right)  \left(  z\right)  =2^{-\frac{1}{2}}z$. If $n=2$ we have
$H_{2}^{\mu}\left(  t\right)  =\frac{4}{1+2\mu}t^{2}-2$ and then $\left(
B_{\mu}f_{2}\right)  \left(  z\right)  =\frac{1}{2}z^{2}+\frac{1+2\mu}{2}$,
and so on. It is clear that $B_{\mu}$ maps polynomials of degree $n$ in
$L^{2}\left(  \mathbb{R},dg_{\mu}\right)  $ to polynomials of degree $n$ in
$\mathcal{B}_{\mu}^{2}$.

Writing $B_{\mu}$ as an integral kernel operator (and, as usual, writing the
kernel also as $B_{\mu}$) we have that
\begin{equation}
\left(  B_{\mu}f\right)  (z)=\int_{\mathbb{R}}B_{\mu}(z,t)f(t)dg_{\mu
}(t),\tag{3.2}%
\end{equation}
where the kernel $B_{\mu}:\mathbb{C}\times\mathbb{R}\rightarrow\mathbb{C}$ is
\[
B_{\mu}(z,t)=\exp\left(  -\frac{z^{2}}{2}\right)  \mathbf{e}_{\mu}\left(
2^{\frac{1}{2}}tz\right)  .
\]

For each $z=x+iy\in\mathbb{C}$ fixed, let us consider the function $t\mapsto
B_{\mu}(z,t)$. If $1<p\leq\infty$ we have that
\begin{align*}
&  \int_{\mathbb{R}}\left\vert B_{\mu}(z,t)\right\vert ^{p^{\prime}}dg_{\mu
}(t)\\
&  =\left(  \Gamma\left(  \mu+\frac{1}{2}\right)  \right)  ^{-1}\left\vert
\exp\left(  -\frac{z^{2}}{2}\right)  \right\vert ^{p^{\prime}}\int
_{\mathbb{R}}\left\vert \mathbf{e}_{\mu}\left(  2^{\frac{1}{2}}tz\right)
\right\vert ^{p^{\prime}}\exp\left(  -t^{2}\right)  \left\vert t\right\vert
^{2\mu}dt\\
&  \leq\left(  \Gamma\left(  \mu+\frac{1}{2}\right)  \right)  ^{-1}\exp\left(
-p^{\prime}\frac{x^{2}-y^{2}}{2}\right)  \int_{\mathbb{R}}\mathbf{e}_{\mu
}\left(  2^{\frac{1}{2}}p^{\prime}tx\right)  \exp\left(  -t^{2}\right)
\left\vert t\right\vert ^{2\mu}dt\\
&  =\exp\left(  \frac{p^{\prime}}{2}\left(  p^{\prime}-1\right)  x^{2}%
+\frac{p^{\prime}}{2}y^{2}\right)  <\infty,
\end{align*}
where we used the inequality (2.2) and the equality
\begin{equation}
\int_{\mathbb{R}}\mathbf{e}_{\mu}\left(  \pm2^{\frac{1}{2}}p^{\prime
}xt\right)  \exp\left(  -t^{2}\right)  \left\vert t\right\vert ^{2\mu
}dt=\Gamma\left(  \mu+\frac{1}{2}\right)  \exp\left(  \frac{p^{\prime2}x^{2}%
}{2}\right)  ,\tag{3.3}%
\end{equation}
which comes from the formula
\[
\int_{\mathbb{R}}\mathbf{e}_{\mu}\left(  -i\tilde{x}t\right)  \mathbf{e}_{\mu
}\left(  i\tilde{y}t\right)  \exp\left(  -\eta t^{2}\right)  \left\vert
t\right\vert ^{2\mu}dt\!=\!\frac{\Gamma\left(  \mu+\frac{1}{2}\right)  }%
{\eta^{\mu+\frac{1}{2}}}\exp\left(  -\frac{\tilde{x}^{2}+\tilde{y}^{2}}{4\eta
}\right)  \mathbf{e}_{\mu}\left(  \frac{\tilde{x}\tilde{y}}{2\eta}\right)
\]
(see [Ros1], p. 379) with $\tilde{x}=\pm i2^{\frac{1}{2}}p^{\prime}x$,
$\tilde{y}=0$ and $\eta=1$. This shows that the function $t\mapsto B_{\mu
}(z,t)$ belongs to the space $L^{p^{\prime}}\left(  \mathbb{R},dg_{\mu
}\right)  $.

Observe that if $f\in L^{p}\left(  \mathbb{R},dg_{\mu}\right)  $,
$1<p\leq\infty$, we have by H\"{o}lder's inequality that
\[
\int_{\mathbb{R}}\left\vert B_{\mu}(z,t)f(t)\right\vert dg_{\mu}(t)\leq\left(
\int_{\mathbb{R}}\left\vert B_{\mu}(z,t)\right\vert ^{p^{\prime}}dg_{\mu
}(t)\right)  ^{\frac{1}{p^{\prime}}}\left(  \int_{\mathbb{R}}\left\vert
f(t)\right\vert ^{p}dg_{\mu}(t)\right)  ^{\frac{1}{p}}<\infty.
\]

That is, $\left(  B_{\mu}f\right)  \left(  z\right)  $ defined in (3.2) makes
sense for any $f\in L^{p}\left(  \mathbb{R},dg_{\mu}\right)  $, $1<p\leq
\infty$ and any $z\in\mathbb{C}$. Observe that Morera's theorem tells us that
$B_{\mu}f:\mathbb{C\rightarrow C}$ is holomorphic. The goal of the next
section will be to identify values of $p\in\left(  1,+\infty\right]  $,
$q\in\left[  1,+\infty\right)  $ and $\lambda>0$ such that $B_{\mu}$ is a
bounded operator from $L^{p}\left(  \mathbb{R},dg_{\mu}\right)  $ to
$\mathcal{B}_{\mu,\lambda}^{q}$. For example, we know that when $p=q=2$,
$\lambda=1$ (and $\mu\geq0$, a situation included in the work of Rosenblum and
Marron), the operator $B_{\mu}$ is bounded, since in this case $B_{\mu} $ is
an isometry. But as we will see in the next section, there are
\textquotedblleft lots\textquotedblright\ of pairs of Lebesgue indices
$(p,q)\in\left(  1,+\infty\right]  \times\left[  1,+\infty\right)  $ (or
equivalently $\left(  p^{-1},q^{-1}\right)  \in\left[  0,1\right)
\times\left(  0,1\right]  $, with the standard conventions $0^{-1}=+\infty$
and $+\infty^{-1}=0$), and values of the parameter $\lambda>0$, for which
$B_{\mu}$ is a bounded operator from $L^{p}\left(  \mathbb{R},dg_{\mu}\right)
$ to $\mathcal{B}_{\mu,\lambda}^{q}$.

What we will do in the next section is to obtain sufficient conditions on the
Lebesgue indices $p$ and $q$, and on the weight $\lambda>0$ for $B_{\mu}$ to
be a Hille-Tamarkin operator from $L^{p}\left(  \mathbb{R},dg_{\mu}\right)  $
to $L^{q}\left(  \mathbb{C}\times\mathbb{Z}_{2},d\nu_{\mu,\lambda}\right)  $.
The rest of this section is devoted to making some observations which will
simplify the work of the proof of Theorem 4.1.

Observe that, for any $f\in L^{p}\left(  \mathbb{R},dg_{\mu}\right)  $ given,
we can write the decomposition of the function $B_{\mu}f$ in its even and odd
parts as
\begin{align*}
\left(  B_{\mu}f\right)  \left(  z\right)   &  =\left(  B_{\mu}f\right)
_{e}\left(  z\right)  +\left(  B_{\mu}f\right)  _{o}\left(  z\right) \\
&  =\int_{\mathbb{R}}B_{e,\mu}(z,t)f(t)dg_{\mu}(t)+\int_{\mathbb{R}}B_{o,\mu
}(z,t)f(t)dg_{\mu}(t),
\end{align*}
where
\[
B_{e,\mu}(z,t) =\frac{1}{2}\exp\left(  -\frac{z^{2}}{2}\right)  \left(
\mathbf{e}_{\mu}\left(  2^{\frac{1}{2}}zt\right)  +\mathbf{e}_{\mu}\left(
-2^{\frac{1}{2}}zt\right)  \right)
\]
and
\[
B_{o,\mu}(z,t) =\frac{1}{2}\exp\left(  -\frac{z^{2}}{2}\right)  \left(
\mathbf{e}_{\mu}\left(  2^{\frac{1}{2}}zt\right)  -\mathbf{e}_{\mu}\left(
-2^{\frac{1}{2}}zt\right)  \right)
\]
are the even and odd parts of $z\mapsto B_{\mu}(z,t)$, respectively. Thus, we
can consider operators $B_{e,\mu}$ and $B_{o,\mu}$ defined for all $f\in
L^{p}\left(  \mathbb{R},dg_{\mu}\right)  $, as $B_{e,\mu}f=\left(  B_{\mu
}f\right)  _{e}$ and $B_{o,\mu}f=\left(  B_{\mu}f\right)  _{o}$, that is
\begin{align*}
\left(  B_{e,\mu}f\right)  (z)  &  =\int_{\mathbb{R}}B_{e,\mu}(z,t)f(t)dg_{\mu
}(t),\\
\left(  B_{o,\mu}f\right)  (z)  &  =\int_{\mathbb{R}}B_{o,\mu}(z,t)f(t)dg_{\mu
}(t).
\end{align*}

So $B_{e,\mu}$ and $B_{o,\mu}$ are integral kernel operators whose kernels are
the even and odd parts of the kernel of the integral kernel operator $B_{\mu}%
$. Suppose that there exist $p\in\left(  1,+\infty\right]  $, $q\in\left[
1,+\infty\right)  $ and $\lambda>0$ such that $B_{\mu}$ is a bounded operator
from $L^{p}\left(  \mathbb{R},dg_{\mu}\right)  $ to $L^{q}\left(
\mathbb{C}\times\mathbb{Z}_{2},d\nu_{\mu,\lambda}\right)  $. (We will see in
the next section that such $p,q,\lambda$ do exist.) Then we have that
$B_{e,\mu}$ is a bounded operator from $L^{p}\left(  \mathbb{R},dg_{\mu
}\right)  $ to $L^{q}\left(  \mathbb{C},d\nu_{e,\mu,\lambda}\right)  $ and
$B_{o,\mu}$ is a bounded operator from $L^{p}\left(  \mathbb{R},dg_{\mu
}\right)  $ to $L^{q}\left(  \mathbb{C},d\nu_{o,\mu,\lambda}\right)  $.
Conversely, if there exist $p\in\left(  1,+\infty\right]  $, $q\in\left[
1,+\infty\right)  $ and $\lambda>0$ such that $B_{e,\mu}$ and $B_{o,\mu}$ are
bounded operators from $L^{p}\left(  \mathbb{R},dg_{\mu}\right)  $ to
$L^{q}\left(  \mathbb{C},d\nu_{e,\mu,\lambda}\right)  $ and to $L^{q}\left(
\mathbb{C},d\nu_{o,\mu,\lambda}\right)  $, respectively, then $B_{\mu}$ is a
bounded operator from $L^{p}\left(  \mathbb{R},dg_{\mu}\right)  $ to
$L^{q}\left(  \mathbb{C}\times\mathbb{Z}_{2},d\nu_{\mu,\lambda}\right)  $.

Finally, let us note that since $B_{\mu}=B_{e,\mu}+B_{o,\mu}$, we have that
$|||B_{\mu}|||_{p,q}\leq|||B_{e,\mu}|||_{p,q}+|||B_{o,\mu}|||_{p,q}$, so if
$B_{e,\mu}$ and $B_{o,\mu}$ are Hille-Tamarkin operators with respect to $p$
and $q$, then $B_{\mu}$ is a Hille-Tamarkin operator with respect to $p$ and
\nolinebreak$q$. \bigskip

\section{$L^{p}$ mapping properties of $B_{\mu}$}

\bigskip The main result in this section is the following.

\bigskip

\textbf{Theorem 4.1 \ }\textit{Let }$1<p\leq\infty$\textit{, }$1\leq q<\infty
$\textit{\ and }$\lambda>\frac{1}{2}$\textit{. A sufficient condition for
}$B_{\mu}$\textit{\ to be a Hille-Tamarkin operator from }$L^{p}\left(
\mathbb{R},dg_{\mu}\right)  $\textit{\ to }$L^{q}\left(  \mathbb{C}%
\times\mathbb{Z}_{2},d\nu_{\mu,\lambda}\right)  $\textit{\ is that }%
$p,q$\textit{\ and }$\lambda$\textit{\ satisfy the inequalities }
\begin{equation}
p>1+\frac{q}{2\lambda}\mathit{\qquad}\text{\textit{and}\qquad}1\leq
q<2\lambda.\tag{4.1}%
\end{equation}

\textit{(Notice that these conditions do not depend on }$\mu$\textit{.)}

\bigskip

\textbf{Remark:} In the case $\mu=0$ these conditions are also necessary for
the operator to be Hille-Tamarkin. See [Snt1]. We conjecture that this is also
true in this more general context.

\bigskip

\textbf{Proof: }As we mentioned in the last section, it is sufficient to prove
that the conditions (4.1) imply that $B_{e,\mu}$ and $B_{o,\mu}$ are
Hille-Tamarkin operators with respect to $p$ and $q$. We begin by considering
$B_{e,\mu}$. We have that
\begin{align*}
\left\vert B_{e,\mu}\left(  z,t\right)  \right\vert  &  =\left\vert \frac
{1}{2}\exp\left(  -\frac{z^{2}}{2}\right)  \left(  \mathbf{e}_{\mu}\left(
2^{\frac{1}{2}}zt\right)  +\mathbf{e}_{\mu}\left(  -2^{\frac{1}{2}}zt\right)
\right)  \right\vert \\
&  \leq\exp\left(  -\frac{x^{2}-y^{2}}{2}\right)  \left(  \left\vert
\mathbf{e}_{\mu}\left(  2^{\frac{1}{2}}zt\right)  \right\vert +\left\vert
\mathbf{e}_{\mu}\left(  -2^{\frac{1}{2}}zt\right)  \right\vert \right)  ,
\end{align*}
where $z=x+iy\in\mathbb{C}$, $x,y\in\mathbb{R}$. Note that this inequality is
also valid for the kernel $B_{o,\mu}\left(  z,t\right)  $. By using (2.2) and
(3.3) we have that
\begin{align*}
&  \left(  \int_{\mathbb{R}}\left\vert B_{e,\mu}\left(  z,t\right)
\right\vert ^{p^{\prime}}dg_{\mu}(t)\right)  ^{\frac{1}{p^{\prime}}}\\
&  \leq\left\{  \int_{\mathbb{R}}\left(  \exp\left(  -\frac{x^{2}-y^{2}}%
{2}\right)  \left(  \left\vert \mathbf{e}_{\mu}\left(  2^{\frac{1}{2}%
}zt\right)  \right\vert +\left\vert \mathbf{e}_{\mu}\left(  -2^{\frac{1}{2}%
}zt\right)  \right\vert \right)  \right)  ^{p^{\prime}}dg_{\mu}(t)\right\}
^{\frac{1}{p^{\prime}}}\\
&  \leq{C}\exp\left(  -\frac{x^{2}-y^{2}}{2}\right)  \left(  \int_{\mathbb{R}%
}\left(  \left\vert \mathbf{e}_{\mu}\left(  2^{\frac{1}{2}}zt\right)
\right\vert ^{p^{\prime}}+\left\vert \mathbf{e}_{\mu}\left(  -2^{\frac{1}{2}%
}zt\right)  \right\vert ^{p^{\prime}}\right)  dg_{\mu}(t)\right)  ^{\frac
{1}{p^{\prime}}}\\
&  \leq{C}\exp\left(  -\frac{x^{2}-y^{2}}{2}\right)  \!\left(  \int
_{\mathbb{R}}\left(  \!\mathbf{e}_{\mu}\!\left(  p^{\prime}2^{\frac{1}{2}%
}xt\right)  +\mathbf{e}_{\mu}\!\left(  -p^{\prime}2^{\frac{1}{2}}xt\right)
\right)  \!\exp\!\left(  -t^{2}\right)  \left\vert t\right\vert ^{2\mu
}\!dt\right)  ^{\frac{1}{p^{\prime}}}\\
&  =C\exp\left(  -\frac{x^{2}-y^{2}}{2}+\frac{p^{\prime}x^{2}}{2}\right)  .
\end{align*}

Thus we obtain
\begin{align*}
{|||B_{e,\mu}|||_{p,q}}=\left(  \int_{\mathbb{C}}\left(  \int_{\mathbb{R}%
}\left\vert B_{e,\mu}\left(  z,t\right)  \right\vert ^{p^{\prime}}dg_{\mu
}(t)\right)  ^{\frac{q}{p^{\prime}}}d\nu_{e,\mu,\lambda}(z)\right)  ^{\frac
{1}{q}}\\
\leq C\left(  \int_{\mathbb{C}}\exp\left(  -q\frac{x^{2}-y^{2}}{2}%
+\frac{qp^{\prime}x^{2}}{2}\right)  K_{\mu-\frac{1}{2}}\left(  \lambda
\left\vert z\right\vert ^{2}\right)  \left\vert z\right\vert ^{2\mu
+1}dxdy\right)  ^{\frac{1}{q}}.
\end{align*}

The last integral is finite if and only if for all $M>0$ we have that
\[
\int_{\left\vert z\right\vert >M}\exp\left(  -q\frac{x^{2}-y^{2}}{2}%
+\frac{qp^{\prime}x^{2}}{2}\right)  K_{\mu-\frac{1}{2}}\left(  \lambda
\left\vert z\right\vert ^{2}\right)  \left\vert z\right\vert ^{2\mu
+1}dxdy<\infty.
\]

But for large enough $M>0$ we can use the asymptotic behavior given in (2.8)
of $K_{\mu-\frac{1}{2}}\left(  \lambda\left\vert z\right\vert ^{2}\right)  $
as $\left\vert z\right\vert \rightarrow\infty$ (which does not depend on the
order of the Macdonald function) to conclude that the last expression is
equivalent to
\[
\int_{\left\vert z\right\vert >M}\exp\left(  \left(  -\frac{q}{2}%
+\frac{qp^{\prime}}{2}-\lambda\right)  x^{2}+\left(  \frac{q}{2}%
-\lambda\right)  y^{2}\right)  \left(  x^{2}+y^{2}\right)  ^{\mu}dxdy<\infty,
\]
which is equivalent to the conditions
\[
-\frac{q}{2}+\frac{qp^{\prime}}{2}-\lambda<0\qquad\text{and\qquad}\frac{q}%
{2}-\lambda<0,
\]
which are the conditions in the hypotheses of the theorem. We have proved that
these conditions guarantee that $B_{e,\mu}$ is a Hille-Tamarkin operator. But,
as we mentioned before, the same estimates obtained for $|||B_{e,\mu}%
|||_{p,q}$ work for $|||B_{o,\mu}|||_{p,q}$, since the Macdonald function
$K_{\mu+\frac{1}{2}}\left(  \lambda\left\vert z\right\vert ^{2}\right)  $ also
has the same asymptotics as $\left\vert z\right\vert \rightarrow\infty$. So
the same conditions guarantee that $B_{o,\mu}$ is a Hille-Tamarkin operator.
So finally we conclude that the conditions on $p,q$ and $\lambda$ in the
theorem imply that $B_{\mu}$ is a Hille-Tamarkin operator from $L^{p}\left(
\mathbb{R},dg_{\mu}\right)  $\ to $L^{q}\left(  \mathbb{C}\times\mathbb{Z}%
_{2},d\nu_{\mu,\lambda}\right)  $, as desired.

\hfill\textbf{Q.E.D.} \bigskip

We have proved that for $p,q$ and $\lambda$ as in (4.1), the Hille-Tamarkin
norm of $B_{\mu}$ is finite, and then Proposition 2.1 allows us to conclude
the boundedness of $B_{\mu}:L^{p}\left(  \mathbb{R},dg_{\mu}\right)
\rightarrow L^{q}\left(  \mathbb{C}\times\mathbb{Z}_{2},d\nu_{\mu,\lambda
}\right)  $. Observe that even though we do not have the case $p=2$, $q=2$ and
$\lambda=1$ included in (4.1), we do have the boundedness of $B_{\mu}$ since
for these values of $p$, $q$, and $\lambda$ the operator $B_{\mu}$ is in fact
unitary. In other words, the conditions imposed by the inequalities (4.1) are
sufficient to conclude the boundedness of $B_{\mu}$, but those conditions are
not necessary. On the other hand, Proposition 2.2 (together with Theorem 4.1)
tells us that the inequalities (4.1) are also sufficient to conclude that
$B_{\mu}$ is a compact operator from $L^{p}\left(  \mathbb{R},dg_{\mu}\right)
$\ to $L^{q}\left(  \mathbb{C}\times\mathbb{Z}_{2},d\nu_{\mu,\lambda}\right)
$. The natural question is if in the case $p=2 $, $q=2 $ and $\lambda=1$ the
operator $B_{\mu}$ is compact. The answer is no. In fact, we know that the
Segal-Bargmann transform $B:L^{2}\left(  \mathbb{R},dg\right)  \rightarrow
\mathcal{B}^{2}$ \textit{is not a compact operator}, since in this case $B$ is
a unitary map onto the\textit{\ infinite-dimensional} space $\mathcal{B}^{2}$.
Thus, even though we have that $\left\Vert B_{\mu}\right\Vert _{2\rightarrow
2}=1$, we have that $|||B_{\mu}|||_{2,2}=\infty$ (again by Proposition 2.2).

The case $\mu=0$ and $\lambda=1$ of Theorem 4.1 is contained in Theorem 3.1 of
[Snt1]. So we have that if $1<p\leq\infty$\textit{, }$1\leq q<\infty$ are such
that the inequalities $p>1+\frac{q}{2}$ and $1\leq q<2$ hold, then the
Segal-Bargmann transform $B:L^{2}\left(  \mathbb{R},dg\right)  \rightarrow
\mathcal{B}^{2}$ is bounded. But in this case we have more: if either
$p<1+\frac{q}{2}$ or $q>2$ holds, the Segal-Bargmann transform $B$ is
unbounded (see Corollary 7.2 in [Snt1]).

The pair $\left(  p^{-1},q^{-1}\right)  \in\left[  0,1\right)  \times\left(
0,1\right]  $ is called \textit{admissible} if $\left\Vert B_{\mu}\right\Vert
_{p\rightarrow q}<\infty$.

The inequalities (4.1) can be written as
\begin{equation}
q^{-1}>\frac{1}{2\lambda}\frac{p^{-1}}{1-p^{-1}}\qquad\text{and\qquad}\frac
{1}{2\lambda}<q^{-1}\leq1.\tag{4.2}%
\end{equation}

In the plane with points $\left(  p^{-1},q^{-1}\right)  $, the curve
\[
q^{-1}=\frac{1}{2\lambda}\frac{p^{-1}}{1-p^{-1}}%
\]
is a hyperbola with vertical asymptote $p^{-1}=1$ and horizontal asymptote
$q^{-1}=-\frac{1}{2\lambda}$. This hyperbola passes through the origin and
intersects the horizontal line $q^{-1}=1$ in $\left(  \frac{2\lambda}%
{2\lambda+1},1\right)  $. Then, if $R$ is the region determined by the
inequalities (4.2), we have $R\cap\left(  \left[  0,1\right)  \times\left(
0,1\right]  \right)  \neq\varnothing$, which shows the existence of a
non-empty region of admissible pairs $\left(  p^{-1},q^{-1}\right)  $ for
which the $\mu$-deformed Segal-Bargmann transform is a bounded operator from
$L^{p}\left(  \mathbb{R},dg_{\mu}\right)  $\ to $L^{q}\left(  \mathbb{C}%
\times\mathbb{Z}_{2},d\nu_{\mu,\lambda}\right)  $. Note that the condition
$\lambda>\frac{1}{2}$ guarantees the existence of $q^{-1}\in\left(
0,1\right]  $ satisfying the inequality $\frac{1}{2\lambda}<q^{-1}\leq1$ of (4.2).

We observe that the fact that $\left(  p^{-1},q^{-1}\right)  $ is an
admissible pair depends on the value of $\lambda$. For example, if
$\lambda=\frac{2}{3}$, the pair $\left(  \frac{1}{4},\frac{4}{5}\right)  $ is
admissible, since for these values of $\lambda$, $p$, and $q$ the inequalities
(4.2) hold. Also, the pair $\left(  \frac{1}{4},\frac{2}{5}\right)  $ is
admissible for $\lambda=2$, but it is not, for example, for $\lambda=1$.
(Certainly one easily checks that for $p^{-1}=\frac{1}{4}$, $q^{-1}=\frac
{2}{5}$, and $\lambda=1$ the inequalities (4.2) do not hold. But as we have
seen before this does not imply that the pair $\left(  \frac{1}{4},\frac{2}%
{5}\right)  $ is not admissible for $\lambda=1$. The conclusion comes from
Corollary 7.2 in [Snt1] mentioned above, since in this case we have $q>2$.) So
we have that if the weight $\lambda$ of the codomain space $L^{q}\left(
\mathbb{C}\times\mathbb{Z}_{2},d\nu_{\mu,\lambda}\right)  $ is fixed and
$\lambda>\frac{1}{2}$, then we always can find pairs $\left(  p^{-1}%
,q^{-1}\right)  $ (those that satisfy (4.2)) for which the $\mu$-deformed
Segal-Bargmann transform $B_{\mu}$ is a bounded operator from $L^{p}\left(
\mathbb{R},dg_{\mu}\right)  $\ to $L^{q}\left(  \mathbb{C}\times\mathbb{Z}%
_{2},d\nu_{\mu,\lambda}\right)  $. Moreover, observe that if we have a fixed
pair $\left(  p^{-1},q^{-1}\right)  $ satisfying the inequalities (4.2) for a
given $\lambda_{1}>\frac{1}{2}$, then these inequalities are also satisfied
for any $\lambda\geq\lambda_{1} $.

But there is another point of view of the situation described above:
\textit{any pair }$\left(  p^{-1},q^{-1}\right)  \in\left[  0,1\right)
\times\left(  0,1\right]  $\textit{\ can be admissible, by taking an adequate
value of }$\lambda$\textit{.} In fact, observe that if we take
\begin{equation}
\lambda> \max\left(  \frac{1}{2q^{-1}},\frac{p^{-1}}{2q^{-1}\left(
1-p^{-1}\right)  }\right)  ,\tag{4.3}%
\end{equation}
then the inequalities (4.2) are satisfied for any $\left(  p^{-1}%
,q^{-1}\right)  \in\left[  0,1\right)  \times\left(  0,1\right]  $. That is,
for any pair $(p,q)\in\left(  1,+\infty\right]  \times\left[  1,+\infty
\right)  $, the $\mu$-deformed Segal-Bargmann transform $B_{\mu}:L^{p}\left(
\mathbb{R},dg_{\mu}\right)  \rightarrow L^{q}\left(  \mathbb{C}\times
\mathbb{Z}_{2},d\nu_{\mu,\lambda}\right)  $, where $\lambda$ is taken as in
(4.3), is a bounded operator.

Figures 1, 2, and 3 show the regions of pairs $\left(  p^{-1},q^{-1}\right)  $
where (4.2) holds in the cases $\lambda=2$, $\lambda=1$ and $\lambda=\frac
{2}{3}$, respectively. So these regions are contained in the regions of
admissible pairs $\left(  p^{-1},q^{-1}\right)  $.

\begin{figure}[pth]
\centerline{\includegraphics[height=5cm]{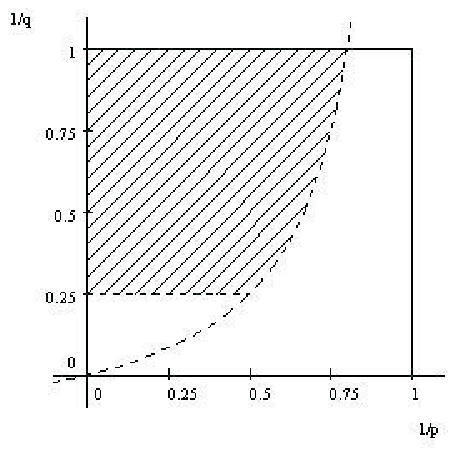}}\caption{{\
Region
where (4.2) holds for }$\lambda=2$: $\frac{1}{4}<q^{-1}\leq1 $, $q^{-1}%
>\frac{p^{-1}}{4\left(  1-p^{-1}\right)  }$.}%
\label{fig:figura1}
\end{figure}

\begin{figure}[pth]
\centerline{\includegraphics[height=5cm]{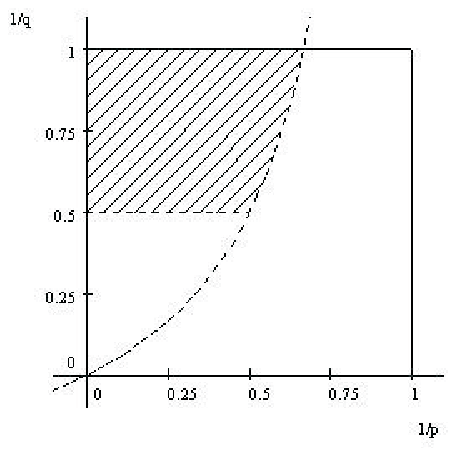}}\caption{{\
Region
where (4.2) holds for }$\lambda=1$: $\frac{1}{2}<q^{-1}\leq1 $, $q^{-1}%
>\frac{p^{-1}}{2\left(  1-p^{-1}\right)  }$.}%
\label{fig:figura2}%
\end{figure}

\begin{figure}[pth]
\centerline{\includegraphics[height=5cm]{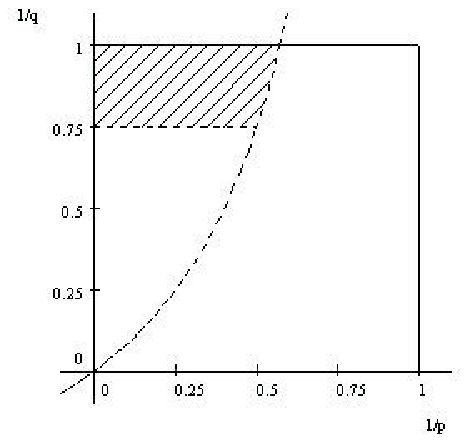}}\caption{{\
Region where (4.2) holds for }$\lambda=\frac{2}{3}$:
$\frac{3}{4}<q^{-1}\leq1 $,
$q^{-1}>\frac{3 p^{-1}}{4\left(  1-p^{-1}\right)  }$.}%
\label{fig:figura3}%
\end{figure}

\bigskip\bigskip

\section{Hirschman inequalities}

\bigskip We know that $B_{\mu}:L^{2}\left(  \mathbb{R},dg_{\mu}\right)
\rightarrow\mathcal{B}_{\mu,\lambda}^{2}$ is a unitary operator when
$\lambda=1$. So the condition $\lambda=1$ is sufficient for $B_{\mu}$ being
unitary. The following result tells us that this condition is also necessary.

\bigskip

\textbf{Proposition 5.1 \ }\textit{Suppose that the operator }$B_{\mu}$
\textit{from }$L^{2}\left(  \mathbb{R},dg_{\mu}\right)  $ \textit{to
}$\mathcal{B}_{\mu,\lambda}^{2}$ \textit{is unitary. Then }$\lambda
=1$\textit{.}

\bigskip

\textbf{Proof: }Let $f$ be a state of $L^{2}\left(  \mathbb{R},dg_{\mu
}\right)  $ (i.e., $\left\Vert f\right\Vert _{L^{2}\left(  \mathbb{R},dg_{\mu
}\right)  }=1$). By taking the orthonormal basis $\left\{  \xi_{n}^{\mu
}\right\}  _{n=0}^{\infty}$ of $\mathcal{B}_{\mu}^{2}$ (see Section 3), we can
write $B_{\mu}f\in\mathcal{B}_{\mu}^{2}$ as $B_{\mu}f=\sum_{n=0}^{\infty}%
a_{n}\xi_{n}^{\mu},$ where the coefficients $a_{n}\in\mathbb{C}$ satisfy
$\sum_{n=0}^{\infty}\left\vert a_{n}\right\vert ^{2}=1$ (since in this case
$B_{\mu}$ is unitary). We can take $f$ such that $a_{k}\neq0$ for some
$k\in\mathbb{N}$. Suppose (in order to get a contradiction) that $\lambda>1$.
By hypothesis we have that $B_{\mu}f\in\mathcal{B}_{\mu,\lambda}^{2}$, so we
can write $B_{\mu}f$ in terms of the basis $\left\{  \chi_{n}^{\mu}\right\}
_{n=0}^{\infty}$ of $\mathcal{B}_{\mu,\lambda}^{2}$ (where $\chi_{n}^{\mu
}=\lambda^{\frac{n}{2}}\xi_{n}^{\mu}$) as $B_{\mu}f=\sum_{n=0}^{\infty}%
\lambda^{-\frac{n}{2}}a_{n}\chi_{n}^{\mu}.$ Since we are assuming that the
operator $B_{\mu}$ from $L^{2}\left(  \mathbb{R},dg_{\mu}\right)  $ to
$\mathcal{B}_{\mu,\lambda}^{2}$ is unitary, we have that $1=\sum_{n=0}%
^{\infty}\left\vert \lambda^{-\frac{n}{2}}a_{n}\right\vert ^{2}=\sum
_{n=0}^{\infty}\lambda^{-n}\left\vert a_{n}\right\vert ^{2}$, and since
$\lambda>1$ and $a_{k}\neq0$ for some $k\in\mathbb{N}$, we have that
$1=\sum_{n=0}^{\infty}\lambda^{-n}\left\vert a_{n}\right\vert ^{2}<\sum
_{n=0}^{\infty}\left\vert a_{n}\right\vert ^{2}=1$, a contradiction. A similar
contradiction occurs in the case $0<\lambda<1$. Thus we conclude that
$\lambda=1$, as desired.

\hfill\textbf{Q.E.D.}

\bigskip

In the same spirit as the Hausdorff-Young inequality (HYI, for short), which
states the boundedness of the Fourier transform $\mathcal{F}:L^{p}\left(
\mathbb{R},dx\right)  \rightarrow L^{p^{\prime}}\left(  \mathbb{R},dx\right)
$ for $p\in\left[  1,2\right]  $ (see [R-S], p. 328), as well as other related
theorems (see [We], pp. 168-9), which also concern boundedness properties of
some operators between $L^{p}$ spaces, we are going to establish an inequality
involving the operator norm of $B_{\mu}$ for a range of values of $p$ and $q$.
This result will play a central role in the demonstration of the main result
of this section. The most important tool used in the proof of this inequality
is the Riesz-Thorin interpolation theorem, which is also used in the
demonstrations of the HYI and the other related theorems mentioned above.

\bigskip\textbf{Theorem 5.1 }(Hausdorff-Young type inequality) \textit{Take
}$1\leq q<2$\textit{, }$p>1+\frac{q}{2}$\textit{\ and }$p_{s}$\textit{\ and
}$q_{s}$\textit{\ defined by }
\[
p_{s}=\left(  sp^{-1}+\left(  1-s\right)  2^{-1}\right)  ^{-1}%
\]
\textit{\ and }
\[
q_{s}=\left(  sq^{-1}+\left(  1-s\right)  2^{-1}\right)  ^{-1}%
\]
\textit{\ for }$s\in\left[  0,1\right]  $\textit{. Then we have }
\[
1\leq\left\Vert B_{\mu}\right\Vert _{p_{s}\rightarrow q_{s}}\leq\left\Vert
B_{\mu}\right\Vert _{p\rightarrow q}^{s}.
\]

\bigskip

\textbf{Proof: }Observe that the pairs $\left(  2^{-1},2^{-1}\right)  $ and
$\left(  p^{-1},q^{-1}\right)  $ are admissible for $B_{\mu}$ and that
$\left\| B_{\mu}\right\| _{{2}\rightarrow{2}}=1$ since $B_{\mu}$ is unitary.
The Riesz-Thorin interpolation theorem (see [B-S], p. 196) says that for any
$s\in\left[  0,1\right]  $, the operator $B_{\mu}$ from $L^{p_{s}}\left(
\mathbb{R},dg_{\mu}\right)  $ to $L^{q_{s}}\left(  \mathbb{C}\times
\mathbb{Z}_{2},d\nu_{\mu}\right)  $ is bounded and that $\left\Vert B_{\mu
}\right\Vert _{p_{s}\rightarrow q_{s}}\leq\left\| B_{\mu}\right\|
_{{p}\rightarrow{q}}^{s}\left\| B_{\mu}\right\| _{{2}\rightarrow{2}}^{1-s}=
\left\Vert B_{\mu}\right\Vert _{p\rightarrow q}^{s}$. Moreover, since $B_{\mu
}1=1$, we have also the inequality $\left\Vert B_{\mu}\right\Vert
_{p_{s}\rightarrow q_{s}}\geq1$, which completes the proof of the theorem.

\hspace{0in}\hfill\textbf{Q.E.D.}

\bigskip

In the main result of this section, which we will present and prove shortly,
we will face the problem of differentiating functions of the form
$\varphi\left(  s\right)  \!=\!\left\Vert f\right\Vert _{L^{T\left(  s\right)
}\left(  \Omega,d\nu\right)  }$ at $s=0$, where $\left(  \Omega,d\nu\right)  $
is a finite measure space, $T:\left[  0,1\right]  \rightarrow\mathbb{R}$ is
the function
\begin{equation}
T\left(  s\right)  =\frac{2\vartheta}{(2-\vartheta)s+\vartheta},\tag{5.1}%
\end{equation}
$\vartheta\geq1$ is a parameter, and $f$ is a non-zero function in the space
$L^{p}\left(  \Omega,d\nu\right)  $ for $p>2$. More precisely, we will need to
calculate the right hand derivative $\varphi^{\prime}\left(  0^{+}\right)  $,
and of course before that, to guarantee its existence.

If we naively calculate $\varphi^{\prime}\left(  0^{+}\right)  $,
interchanging when necessary the differentiation with integration and applying
the rules from elementary calculus, we get
\begin{equation}
\varphi^{\prime}\left(  0^{+}\right)  =\left(  \frac{1}{2}-\frac{1}{\vartheta
}\right)  \left\Vert f\right\Vert _{L^{2}\left(  \Omega,d\nu\right)  }%
^{-1}S_{L^{2}\left(  \Omega,d\nu\right)  }\left(  f\right)  ,\tag{5.2}%
\end{equation}
where $S_{L^{2}\left(  \Omega,d\nu\right)  }\left(  f\right)  $ is the entropy
of $f$, defined in (2.11).

Note that by the very definition of $\varphi^{\prime}\left(  0^{+}\right)  $ a
\textit{necessary condition} for the existence of this derivative is that
$\varphi\left(  s\right)  $ be finite in some interval of the form $\left[
0,\varepsilon\right)  $. That is, we need that the function $f$
\textit{belong} to $L^{T\left(  s\right)  }\left(  \Omega,d\nu\right)  $ for
$0\leq s<\varepsilon$. Let us write this necessary condition as (NC). We can
guarantee (NC), if for example we require that $f\in L^{2+\zeta}\left(
\Omega,d\nu\right)  $ where $\zeta>0$, since in this case we have $T\left(
0\right)  =2<2+\zeta$ which implies that there exists $\varepsilon>0$ such
that $T\left(  s\right)  <2+\zeta$ for $0\leq s<\varepsilon$ which in turn
implies (using H\"{o}lder's inequality) that $\left\Vert f\right\Vert
_{L^{T\left(  s\right)  }\left(  \Omega,d\nu\right)  }\leq C\left\Vert
f\right\Vert _{L^{2+\zeta}\left(  \Omega,d\nu\right)  }<\infty$ for $0\leq
s<\varepsilon$. That is, the condition $f\in L^{2+\zeta}\left(  \Omega
,d\nu\right)  $ where $\zeta>0$ is a \textit{sufficient condition} (denoted
(SC)) for (NC). (We mention that (SC) is not necessary for (NC), since if
$1\leq\vartheta<2$ we have that $f\in L^{2}\left(  \Omega,d\nu\right)  $ is
enough to imply (NC) as one can easily check.) Surprisingly, the condition
(SC) is also a \textit{sufficient condition} for the existence of
$\varphi^{\prime}\left(  0^{+}\right)  $, and in such a case the formula (5.2)
obtained by formal derivation is the \textit{right} formula for this
derivative. This is what the following lemma says; it is Lemma 1.1 of [G] with
some minor changes.

\bigskip

\textbf{Lemma 5.1 }\textit{Let }$\left(  \Omega,d\nu\right)  $\textit{\ be a
finite measure space. Suppose }$\varepsilon>0$\textit{, }$1<p_{0}<\infty
$\textit{, and }$p>p_{0}$\textit{. Let }$T\left(  s\right)  $\textit{\ be a
real continuously differentiable function on }$\left[  0,\varepsilon\right)
$\textit{\ such that }$T\left(  0\right)  =p_{0}$\textit{, and let }$F\left(
s\right)  $\textit{\ be a continuously differentiable function on }$\left[
0,\varepsilon\right)  $\textit{\ into }$L^{p}\left(  \Omega,\nu\right)
$\textit{\ with }$F\left(  0\right)  =f\neq0$\textit{. Then }$\left\Vert
F\left(  s\right)  \right\Vert _{L^{T\left(  s\right)  }\left(  \Omega
,d\nu\right)  }$\textit{\ is differentiable from the right at }$s=0$%
\textit{\ and its derivative is given by}
\begin{align}
&  \left.  \frac{d}{ds}\right\vert _{s=0^{+}}\left\Vert F\left(  s\right)
\right\Vert _{L^{T\left(  s\right)  }\left(  \Omega,d\nu\right)  }\tag{5.3}\\
&  =\!N^{1-p_{0}}\!\left(  p_{0}^{-1}T^{\prime}\!\!\left(  0^{+}\right)
\!\!\left(  \int_{\Omega}\left\vert f\right\vert ^{p_{0}}\log\left\vert
f\right\vert d\nu-N^{p_{0}}\log N\right)  \!+\!\operatorname{Re}\left\langle
F^{\prime}\left(  0^{+}\right)  ,f_{p_{0}}\right\rangle \right)  ,\nonumber
\end{align}
\textit{where} $N$=$\left\Vert f\right\Vert _{L^{p_{0}}\left(  \Omega
,d\nu\right)  }$ \textit{and } $f_{p_{0}}=\left(  \operatorname*{sgn}f\right)
\left\vert f\right\vert ^{p_{0}-1}$\textit{.}\newline\textit{We emphasize that
under these hypotheses, the derivative (5.3) is a finite real number.}

\bigskip

The sign of $z\in\mathbb{C}$, denoted by $\operatorname*{sgn}z$, is defined as
$\operatorname*{sgn}z=z/{\left\vert z\right\vert }$ if $z\neq0$, and
$\operatorname*{sgn}z=0$ if $z=0$. In the case we are dealing with, namely
$\varphi\left(  s\right)  =\left\Vert f\right\Vert _{L^{T\left(  s\right)
}\left(  \Omega,d\nu\right)  }$, we have $p_{0}=2$, $p=2+\zeta$ with $\zeta
>0$, $T\left(  s\right)  $ given by (5.1) (so that $T^{\prime}\left(
0\right)  =-\frac{2}{\vartheta}\left(  2-\vartheta\right)  $), and $F$ is
constant (equal to $f$ for all $s$, so that $F^{\prime}\left(  0\right)  =0$).
Thus, if we denote the norm $\left\Vert f\right\Vert _{L^{2}\left(
\Omega,d\nu\right)  }$ by $N $, the formula (5.3) is in our case%

\begin{align*}
\left.  \frac{d}{ds}\right\vert _{s=0^{+}}\!\!\left\Vert f\right\Vert
_{L^{T\left(  s\right)  }\left(  \Omega,d\nu\right)  }\!\! &  =\!N^{-1}%
2^{-1}\!\left(  \frac{-{2}\left(  2-\vartheta\right)  }{\vartheta}\right)
\!\left(  \int_{\Omega}\left\vert f\right\vert ^{2}\log\left\vert f\right\vert
d\nu\!-\!N^{2}\log N\!\right) \\
&  =\left(  \frac{1}{2}-\frac{1}{\vartheta}\right)  N^{-1}\left(  \int
_{\Omega}\left\vert f\right\vert ^{2}\log\left\vert f\right\vert ^{2}%
d\nu-N^{2}\log N^{2}\right) \\
&  =\left(  \frac{1}{2}-\frac{1}{\vartheta}\right)  N^{-1}S_{L^{2}\left(
\Omega,d\nu\right)  }\left(  f\right)  ,
\end{align*}
which is the formula (5.2) for $\varphi^{\prime}\left(  0^{+}\right)  $.

Roughly speaking, an \textit{uncertainty principle} is an inequality involving
the variance of a function $f$ and the variance of its Fourier transform
$\mathcal{F}f$. (See [Fol], p. 27, for a more general statement of an
uncertainty principle.) For example, the Heisenberg uncertainty principle
states that for any $f\in L^{2}\left(  \mathbb{R},dx\right)  $ such that
$\left\|  f\right\|  _{L^{2}\left(  \mathbb{R},dx\right)  }=1$ one has
\[
\left[  \left(  \int_{\mathbb{R}}\left(  x-\Lambda\right)  ^{2}\left\vert
f\left(  x\right)  \right\vert ^{2}dx\right)  ^{\frac{1}{2}}\right]  \left[
\left(  \int_{\mathbb{R}}\left(  x-\widehat{\Lambda}\right)  ^{2}\left\vert
\left(  \mathcal{F}f\right)  \left(  x\right)  \right\vert ^{2}dx\right)
^{\frac{1}{2}}\right]  \geq\left(  4\pi\right)  ^{-1},
\]
where the factors on the left hand side are the variances of $f$ and of
$\mathcal{F}f$ and
\[
\Lambda=\int_{\mathbb{R}}x\left\vert f\left(  x\right)  \right\vert
^{2}dx\text{ \ \ \ and \ \ }\widehat{\Lambda}=\int_{\mathbb{R}}x\left\vert
\left(  \mathcal{F}f\right)  \left(  x\right)  \right\vert ^{2}dx
\]
(assumed to be finite) are the expected values of $f$ and $\mathcal{F}f$,
respectively. What this inequality tells us is that the variances of $f$ and
$\mathcal{F}f$ can not be simultaneously arbitrarily small. Of course, this
has to do with the well known physical version of the Heisenberg uncertainty
principle about the impossibility of determining simultaneously position and
momentum of a quantum particle.

In his paper [Hir], Hirschman obtained an inequality involving not the
variances of $f$ and $\mathcal{F}f$, but their entropies. Specifically, he
showed that for $f\in L^{2}\left(  \mathbb{R},dx\right)  $ such that
$\left\Vert f\right\Vert _{L^{2}\left(  \mathbb{R},dx\right)  }=1$ one has
\[
S_{L^{2}\left(  \mathbb{R},dx\right)  }\left(  f\right)  +S_{L^{2}\left(
\mathbb{R},dx\right)  }\left(  \mathcal{F}f\right)  \leq0,
\]
whenever the left hand side has meaning. Note that $\left(  \mathbb{R}%
,dx\right)  $ is not a finite measure space, so that one or both of the terms
in the left hand side can be meaningless. In fact, Hirschman conjectured a
sharper upper bound, namely ~$\log2-1$, for the left hand side of the previous
inequality. However, Beckner in [Be] proved this conjecture. The idea behind
Hirschman's method for proving the inequality above is to view each side of
the HYI $\left\|  \mathcal{F}f\right\|  _{p^{\prime}}\leq\left\|  f\right\|
_{p}$ which is valid for $p\in\left[  1,2\right]  $, as a function of $p$ for
fixed $f\in L^{2}\left(  \mathbb{R},dx\right)  $. It turns out that these
functions are smooth, and then it makes sense to take the derivative at
$p=2^{-}$ in both sides of the inequality. The point is that, when $p=2$, the
HYI is in fact an equality, by Plancherel's theorem, and so the derivative
$\left.  \frac{d}{dp}\right|  _{p=2^{-}}$ acts as an order-reversing operator
giving in this way a new inequality, ``the differentiated HYI at $p=2^{-}$''.
It turns out that this yields Hirschman's result. All these ideas were applied
in the context of Segal-Bargmann analysis by the second author ([Snt1]) in the
case $\mu=0$. Following the same kind of ideas, we now establish the main
result of this section.

\bigskip

\textbf{Theorem 5.2 }(Hirschman inequality) \textit{Suppose that }%
$p$\textit{\ and }$q$ \textit{satisfy}
\[
1\leq q<2\qquad\text{\textit{and}\qquad}p>1+\frac{q}{2}.
\]
\textit{Let }$f\in L^{2+\zeta}\left(  \mathbb{R},dg_{\mu}\right)
$\textit{\ with }$\zeta>0$\textit{\ be such that }$B_{\mu}f\in L^{2+\xi
}\left(  \mathbb{C}\times\mathbb{Z}_{2},d\nu_{\mu}\right)  $\textit{\ for some
}$\xi>0$\textit{. Then the Hirschman entropy inequality}
\begin{align}
&  \left(  p^{-1}-2^{-1}\right)  S_{L^{2}\left(  \mathbb{R},dg_{\mu}\right)
}\left(  f\right) \tag{5.4}\\
&  \leq\left(  q^{-1}-2^{-1}\right)  S_{L^{2}\left(  \mathbb{C}\times
\mathbb{Z}_{2},d\nu_{\mu}\right)  }\left(  B_{\mu}f\right)  +\left(
\log\left\Vert B_{\mu}\right\Vert _{p\rightarrow q}\right)  \left\Vert
f\right\Vert _{L^{2}\left(  \mathbb{R},dg_{\mu}\right)  }^{2}\nonumber
\end{align}
\textit{holds}.

\bigskip

\textbf{Remark:} We comment that the set of functions for which the hypotheses
of Theorem 5.2 hold is a dense subspace of $L^{2} ( \mathbb{R} , dg_{\mu})$.
This is shown in the Remark after Theorem 6.3.

\bigskip

\textbf{Proof: }We first note that if $f=0$ the inequality to prove is
trivial, both sides of it being equal to zero. So we take an arbitrary $f$
satisfying the hypotheses with $f\neq0$. Observe that the coefficient of the
norm term in (5.4) is non-negative, since $\left\Vert B_{\mu}\right\Vert
_{p\rightarrow q}\geq1$. So the term itself is non-negative. Nevertheless, the
remaining two terms (the entropy terms) can be positive, negative or zero. In
fact, even though $S_{L^{2}\left(  \mathbb{R},dg_{\mu}\right)  }\left(
f\right)  \geq0$ (since $\left(  \mathbb{R},dg_{\mu}\right)  $ is a
probability measure space), the hypotheses allow the coefficient $\left(
p^{-1}-2^{-1}\right)  $ to be positive, negative or zero. Also, the hypotheses
give us that $\left(  q^{-1}-2^{-1}\right)  >0$, but the entropy
$S_{L^{2}\left(  \mathbb{C}\times\mathbb{Z}_{2},d\nu_{\mu}\right)  }\left(
B_{\mu}f\right)  $ can be positive, negative or zero. (Recall that $\left(
\mathbb{C}\times\mathbb{Z}_{2},d\nu_{\mu}\right)  $ is a measure space with
weight strictly greater than $1$.)

The idea of the proof consists in considering the Hausdorff-Young type
inequality $\left\Vert B_{\mu}\right\Vert _{p_{s}\rightarrow q_{s}}%
\leq\left\Vert B_{\mu}\right\Vert _{p\rightarrow q}^{s}$ we proved above
(Theorem 5.1), where $p_{s}=\left(  sp^{-1}+\left(  1-s\right)  2^{-1}\right)
^{-1}$ and $q_{s}=\left(  sq^{-1}+\left(  1-s\right)  2^{-1}\right)  ^{-1}$,
with $s\in\left[  0,1\right]  $. Observe that these formulas for $p_{s}$ and
$q_{s}$ are of the form $T\left(  s\right)  =\frac{2\vartheta}{(2-\vartheta
)s+\vartheta}$, with $\vartheta=p$ and $\vartheta=q$, respectively, as in the
discussion previous to the theorem. That is, we begin by considering the
inequality $\left\Vert B_{\mu}f\right\Vert _{L^{q_{s}}\left(  \mathbb{C}%
\times\mathbb{Z}_{2},d\nu_{\mu}\right)  }\leq A^{s}\left\Vert f\right\Vert
_{L^{p_{s}}\left(  \mathbb{R},dg_{\mu}\right)  }$, where $A=\left\Vert B_{\mu
}\right\Vert _{p\rightarrow q}$. The point here is to notice that when $s=0$,
this inequality is, in fact, an equality (since the operator $B_{\mu}$ from
$L^{2}\left(  \mathbb{R},dg_{\mu}\right)  $ to $\mathcal{B}_{\mu}^{2}$ is
unitary). Then, by differentiating both sides of it at $s=0^{+}$, we get a new
inequality
\[
\left.  \frac{d}{ds}\right\vert _{s=0^{+}}\left\Vert B_{\mu}f\right\Vert
_{L^{q_{s}}\left(  \mathbb{C}\times\mathbb{Z}_{2},d\nu_{\mu}\right)  }%
\leq\left.  \frac{d}{ds}\right\vert _{s=0^{+}}\left(  A^{s}\left\Vert
f\right\Vert _{L^{p_{s}}\left(  \mathbb{R},dg_{\mu}\right)  }\right)
\]
or
\begin{equation}
\left.  \frac{d}{ds}\right\vert _{s=0^{+}}\left\Vert B_{\mu}f\right\Vert
_{L^{q_{s}}\left(  \mathbb{C}\times\mathbb{Z}_{2},d\nu_{\mu}\right)  }%
\leq\left(  \log A\right)  \left\Vert f\right\Vert _{L^{2}\left(
\mathbb{R},dg_{\mu}\right)  }+\left.  \frac{d}{ds}\right\vert _{s=0^{+}%
}\left\Vert f\right\Vert _{L^{p_{s}}\left(  \mathbb{R},dg_{\mu}\right)
}.\tag{5.5}%
\end{equation}

Note that according to Lemma 5.1, the hypotheses on $f$ and on $B_{\mu}f$
guarantee the existence of the derivatives in this expression. Then we can use
formula (5.2) to obtain
\[
\left.  \frac{d}{ds}\right\vert _{s=0^{+}}\left\Vert f\right\Vert _{L^{p_{s}%
}\left(  \mathbb{R},dg_{\mu}\right)  }=\left(  2^{-1}-p^{-1}\right)
\left\Vert f\right\Vert _{L^{2}\left(  \mathbb{R},dg_{\mu}\right)  }%
^{-1}S_{L^{2}\left(  \mathbb{R},dg_{\mu}\right)  }\left(  f\right)
\]
and
\begin{align}
&  \left.  \frac{d}{ds}\right\vert _{s=0^{+}}\left\Vert B_{\mu}f\right\Vert
_{L^{q_{s}}\left(  \mathbb{C}\times\mathbb{Z}_{2},d\nu_{\mu}\right)
}\nonumber\\
&  =\left(  2^{-1}-q^{-1}\right)  \left\Vert B_{\mu}f\right\Vert _{L^{q_{s}%
}\left(  \mathbb{C}\times\mathbb{Z}_{2},d\nu_{\mu}\right)  }^{-1}%
S_{L^{2}\left(  \mathbb{C}\times\mathbb{Z}_{2},d\nu_{\mu}\right)  }\left(
B_{\mu}f\right)  .\nonumber
\end{align}

Thus, inequality (5.5) becomes
\begin{align*}
&  \left(  2^{-1}-q^{-1}\right)  \left\Vert B_{\mu}f\right\Vert _{L^{2}\left(
\mathbb{C}\times\mathbb{Z}_{2},d\nu_{\mu}\right)  }^{-1}S_{L^{2}\left(
\mathbb{C}\times\mathbb{Z}_{2},d\nu_{\mu}\right)  }\left(  B_{\mu}f\right) \\
&  \leq\left(  \log A\right)  \left\Vert f\right\Vert _{L^{2}\left(
\mathbb{R},dg_{\mu}\right)  }+\left(  2^{-1}-p^{-1}\right)  \left\Vert
f\right\Vert _{L^{2}\left(  \mathbb{R},dg_{\mu}\right)  }^{-1}S_{L^{2}\left(
\mathbb{R},dg_{\mu}\right)  }\left(  f\right)  ,
\end{align*}
and finally, by using the fact $\left\Vert B_{\mu}f\right\Vert _{L^{2}\left(
\mathbb{C}\times\mathbb{Z}_{2},d\nu_{\mu}\right)  }=\left\Vert f\right\Vert
_{L^{2}\left(  \mathbb{R},dg_{\mu}\right)  }$, we obtain the inequality (5.4).

\hspace{0in}\hfill\textbf{Q.E.D.}

\bigskip

\textbf{Remark:} This proof depends on the fact that $B_{\mu}$ is a unitary
operator for $p=q=2$ and $\lambda=1$. We can not extend this proof to the case
$p=q=2$ and $\lambda\neq1$ by Proposition 5.1 \bigskip

\section{Logarithmic Sobolev Inequalities}

\bigskip Throughout this section the parameter $\lambda\geq1$ will be assumed.

The term ``Sobolev inequality'' refers to an estimate of lower order
derivatives of a function in terms of its higher order derivatives. Ever since
the work of Sobolev ([Sob]), this kind of estimate has proven to be very
useful in the theory of partial differential equations. (See [L-L], chapter
8.) An example of a Sobolev inequality for a function $f:\mathbb{R}%
^{n}\rightarrow\mathbb{C}$ is
\[
S_{n}\left\|  f\right\|  _{L^{q}\left(  \mathbb{R}^{n},dx\right)  }^{2}%
\leq\left\|  \operatorname*{grad}f\right\|  _{L^{2}\left(  \mathbb{R}%
^{n},dx\right)  }^{2},
\]
where $n\geq3$, $q=2n\left(  n-2\right)  ^{-1}$ and $S_{n}$ a universal
constant depending only on $n$. (See [L-L], p. 156.)

In 1975, Gross ([G]) obtained the inequality
\begin{align}
&  \int_{\mathbb{R}^{n}}\left\vert f\left(  x\right)  \right\vert ^{2}%
\log\left\vert f\left(  x\right)  \right\vert d\nu\left(  x\right)
-\left\Vert f\right\Vert _{L^{2}\left(  \mathbb{R}^{n},d\nu\right)  }^{2}%
\log\left\Vert f\right\Vert _{L^{2}\left(  \mathbb{R}^{n},d\nu\right)
}\nonumber\\
&  \leq\int_{\mathbb{R}^{n}}\left\vert \operatorname*{grad}f\left(  x\right)
\right\vert ^{2}d\nu\left(  x\right)  ,\nonumber
\end{align}
valid for suitable functions $f:\mathbb{R}^{n}\rightarrow\mathbb{C}$, where
$d\nu$ is a Gaussian measure on $\mathbb{R}^{n}$. This inequality has the same
flavor of the Sobolev inequality mentioned above, since on both right hand
sides appears the $L^{2}$ norm of $\operatorname*{grad}f$, and on the left
hand side appears an $L^{p}$ norm of the function itself, with some mixed
log's in the latter case. Gross refers to this result as a \textit{logarithmic
Sobolev inequality}, and this type of inequality has been shown since Gross'
work in a variety of generalizations. In particular, in [Snt1] a logarithmic
Sobolev inequality (LSI, for short) is obtained in the context of
Segal-Bargmann analysis. Following the same sort of ideas, we will obtain in
this section a LSI for the $\mu$-deformed Segal-Bargmann space and its
associated transform.

Recall that the two main steps in the development of the theory in the last
section were first to have a Hausdorff-Young type inequality (in order to have
an inequality between operator norms that are smooth functions of the
corresponding Lebesgue indices), and second to use this inequality in order to
obtain the Hirschman inequality (by applying the differentiation technique of
Hirschman to the inequality of the first step). We will follow in this section
the analogues of these steps by first proving another Hausdorff-Young type
inequality and then using this inequality to obtain the LSI desired.

Instead of the Riesz-Thorin interpolation theorem we used to prove the
Hausdorff-Young type inequality in the previous section, we will use here a
generalization of it (Stein's theorem) which we quote next. Recall that a
\textit{simple function} is a measurable function $f$ having a finite range
$R\subset\mathbb{C}$ such that $f^{-1}\left(  z\right)  $ is a set of finite
measure for every $z\in R$, $z\neq0$.

\bigskip

\textbf{Theorem 6.1 }(Stein [St]) \textit{Let }$\left(  \Omega_{j},d\nu
_{j}\right)  $\textit{\ for }$j=1,2$\textit{\ be }$\sigma$-\textit{finite
measure spaces. Let }$T$\textit{\ be a linear transformation which takes
simple functions }$f:\Omega_{1}\rightarrow\mathbb{C}$\textit{\ to measurable
functions }$Tf:\Omega_{2}\rightarrow\mathbb{C}$\textit{. Let }$p_{i},q_{i}%
\in\left[  1,\infty\right]  $\textit{, }$i=0,1$\textit{. Then, for }%
$s\in\left[  0,1\right]  $\textit{, define }$p_{s} $\textit{\ and } $q_{s}%
$\textit{\ by} $p_{s}^{-1}=\left(  1-s\right)  p_{0}^{-1}+sp_{1}^{-1}$
\textit{\ and} $q_{s}^{-1}=\left(  1-s\right)  q_{0}^{-1}+sq_{1}^{-1}.$
\textit{\newline} \textit{For }$i=0,1$\textit{, suppose that }$u_{i}%
:\Omega_{1}\rightarrow\left[  0,\infty\right)  $\textit{\ and }$k_{i}%
:\Omega_{2}\rightarrow\left[  0,\infty\right)  $\textit{\ are measurable
functions with the property that for all simple functions }$f:\Omega
_{1}\rightarrow\mathbb{C}$\textit{\ there exist finite non-negative constants
}$A_{i}$\textit{\ such that}
\begin{equation}
\left\Vert \left(  Tf\right)  k_{i}\right\Vert _{L^{q_{i}}\left(  \Omega
_{2},d\nu_{2}\right)  }\leq A_{i}\left\Vert fu_{i}\right\Vert _{L^{p_{i}%
}\left(  \Omega_{1},d\nu_{1}\right)  }\mathit{.}\tag{6.1}%
\end{equation}

\textit{For }$s\in\left[  0,1\right]  $\textit{, define functions }%
$u_{s}:\Omega_{1}\rightarrow\left[  0,\infty\right)  $ \textit{and }%
$k_{s}:\Omega_{2}\rightarrow\left[  0,\infty\right)  $\textit{, by}
$u_{s}=u_{0}^{1-s}u_{1}^{s}$\textit{\ and }$k_{s}=k_{0}^{1-s}k_{1}^{s}%
$.\textit{\ Then the transformation }$T$\textit{\ can be extended uniquely to
a linear transformation defined on the space of all }$f:\Omega_{1}%
\rightarrow\mathbb{C}$\textit{\ that satisfy }$\left\Vert fu_{s}\right\Vert
_{L^{p_{s}}\left(  \Omega_{1},d\nu_{1}\right)  }<\infty$\textit{\ in such a
way that for all such }$f$\textit{\ we have}
\begin{equation}
\left\Vert \left(  Tf\right)  k_{s}\right\Vert _{L^{q_{s}}\left(  \Omega
_{2},d\nu_{2}\right)  }\leq A_{0}^{1-s}A_{1}^{s}\left\Vert fu_{s}\right\Vert
_{L^{p_{s}}\left(  \Omega_{1},d\nu_{1}\right)  }.\tag{6.2}%
\end{equation}

\bigskip

We will need later the following result.

\bigskip\textbf{Lemma 6.1 }\textit{Let }$1\leq q<2\lambda$\textit{\ and
}$0\leq s\leq1$.\textit{\ Let the function }$\kappa_{\lambda,s}:\mathbb{C}%
\times\mathbb{Z}_{2}\rightarrow\left[  0,\infty\right)  $\textit{\ be defined
by}
\begin{align*}
\kappa_{\lambda,s}\left(  z,1\right)   &  =\left(  \frac{\lambda^{\frac
{2\mu+3}{2}}K_{\mu-\frac{1}{2}}\left(  \lambda\left\vert z\right\vert
^{2}\right)  }{K_{\mu-\frac{1}{2}}\left(  \left\vert z\right\vert ^{2}\right)
}\right)  ^{sq^{-1}},\\
\kappa_{\lambda,s}\left(  z,-1\right)   &  =\left(  \frac{\lambda^{\frac
{2\mu+3}{2}}K_{\mu+\frac{1}{2}}\left(  \lambda\left\vert z\right\vert
^{2}\right)  }{K_{\mu+\frac{1}{2}}\left(  \left\vert z\right\vert ^{2}\right)
}\right)  ^{sq^{-1}}.
\end{align*}
\textit{Then }$\kappa_{\lambda,s}\in L^{\infty}\left(  \mathbb{C}%
\times\mathbb{Z}_{2}\right)  $.

\bigskip

\textbf{Proof: \ }We will prove that the restrictions of $\kappa_{\lambda,s}$
to each copy of $\mathbb{C}$ in $\mathbb{C}\times\mathbb{Z}_{2}$ are bounded
functions in a neighborhood of the origin and in a neighborhood of infinity,
from which the conclusion of the lemma follows. We begin by considering
$\kappa_{\lambda,s}\left(  z,1\right)  $ in a neighborhood of $\left(
0,1\right)  $. By applying (2.6) we find that if $0\leq\mu<\frac{1}{2}$ we
have that $\kappa_{\lambda,s}\left(  z,1\right)  \cong\lambda^{\frac{\left(
2\mu+1\right)  s}{q}}$ as $\left\vert z\right\vert \rightarrow0$, and if
$\mu>\frac{1}{2}$ we have that $\kappa_{\lambda,s}\left(  z,1\right)
\cong\lambda^{\frac{2s}{q}}$ as $\left\vert z\right\vert \rightarrow0$. This
shows that for $\mu\neq\frac{1}{2}$, the function $\kappa_{\lambda,s}\left(
z,1\right)  $ is bounded in a neighborhood of $\left(  0,1\right)  $. In the
case $\mu=\frac{1}{2}$ we have by (2.7) that
\[
\kappa_{\lambda,s}\left(  z,1\right)  \cong\left(  \frac{\lambda^{2}\log
\frac{2}{\lambda\left\vert z\right\vert ^{2}}}{\log\frac{2}{\left\vert
z\right\vert ^{2}}}\right)  ^{sq^{-1}}.
\]

But the right hand side of this expression is bounded in a neighborhood of the
origin since it has the finite limit $\lambda^{\frac{2s}{q}}$ as $\left\vert
z\right\vert \rightarrow0$. Again using (2.6) we have that $\kappa_{\lambda
,s}\left(  z,-1\right)  \cong\lambda^{\frac{s}{q}}$ as $\left\vert
z\right\vert \rightarrow0$ for all $\mu\geq0$, which shows that $\kappa
_{\lambda,s}\left(  z,-1\right)  $ is bounded near $\left(  0,-1\right)  $.

Finally, according to (2.8) we have that both $\kappa_{\lambda,s}\left(
z,1\right)  $ and $\kappa_{\lambda,s}\left(  z,-1\right)  $ are asymptotically
equivalent as $\left\vert z\right\vert \rightarrow+\infty$ to
\[
\left(  \frac{\lambda^{\frac{2\mu+3}{2}}\left(  \frac{\pi}{2\lambda}\right)
^{\frac{1}{2}}\left\vert z\right\vert ^{-1}\exp\left(  -\lambda\left\vert
z\right\vert ^{2}\right)  }{\left(  \frac{\pi}{2}\right)  ^{\frac{1}{2}%
}\left\vert z\right\vert ^{-1}\exp\left(  -\left\vert z\right\vert
^{2}\right)  }\right)  ^{sq^{-1}}=\lambda^{\frac{\left(  \mu+1\right)  s}{q}%
}\exp\left(  \frac{1-\lambda}{q}s\left\vert z\right\vert ^{2}\right)  ,
\]
which is a bounded function of $z$, since $\lambda\geq1$.

\hfill\textbf{Q.E.D.}

\bigskip

We will prove now a Hausdorff-Young type inequality as we did in Theorem 5.1.
Recall that in Section 5 we worked with the operator $B_{\mu}$ from
$L^{p}\left(  \mathbb{R},dg_{\mu}\right)  $ to $L^{q}\left(  \mathbb{C}%
\times\mathbb{Z}_{2},d\nu_{\mu}\right)  $ with $p$ and $q$ chosen in such a
way that $\Pi_{1}=\left(  p^{-1},q^{-1}\right)  $ is admissible. Then we used
the Riesz-Thorin interpolation theorem to conclude that for all pairs $\left(
p_{s}^{-1},q_{s}^{-1}\right)  =s\Pi_{1}+\left(  1-s\right)  \Pi_{2}$, $0\leq
s\leq1$, in the line segment connecting $\Pi_{2}=\left(  2^{-1},2^{-1}\right)
$ and $\Pi_{1}$, the corresponding operator $B_{\mu}$ from $L^{p_{s}}\left(
\mathbb{R},dg_{\mu}\right)  $ to $L^{q_{s}}\left(  \mathbb{C}\times
\mathbb{Z}_{2},d\nu_{\mu}\right)  $ is bounded and that $1\leq\left\|  B_{\mu
}\right\|  _{p_{s}\rightarrow q_{s}}\leq\left\|  B_{\mu}\right\|
_{p\rightarrow q}^{s}$ for all $s\in\left[  0,1\right]  $. Since the operator
$B_{\mu}$ from $L^{2}\left(  \mathbb{R},dg_{\mu}\right)  $ to $L^{2}\left(
\mathbb{C}\times\mathbb{Z}_{2},d\nu_{\mu}\right)  $ is isometric we have that
$\Pi_{2}$ is admissible. Notice that the measure $d\nu_{\mu}$ of the spaces
$L^{q_{s}}\left(  \mathbb{C}\times\mathbb{Z}_{2},d\nu_{\mu}\right)  $ is
independent of the parameter $s\in\left[  0,1\right]  $. What we will do now
will be something like repeating this story in another setting, using the
Stein's interpolation theorem instead of the Riesz-Thorin theorem, in such a
way that we get the same sort of result: an inequality for the operator norm
of the operators from the $L^{p_{s}}$ spaces to the $L^{q_{s}}$ spaces, such
that when $s=0$ this inequality becomes an equality. The price to be paid has
to do with the measure of the $L^{q_{s}}$ codomain spaces, which now will
depend on the parameter $s$. The result is the following.

\bigskip

\textbf{Theorem 6.2 \ }(Weighted Hausdorff-Young type inequality)
\ \textit{Let }$p,q,\lambda$\textit{\ be parameters as in (4.1). For }%
$s\in\left[  0,1\right]  $ \textit{let }$\kappa_{\lambda,s}:\mathbb{C}%
\times\mathbb{Z}_{2}\rightarrow\left[  0,\infty\right)  $ \textit{be the
function defined in Lemma 6.1, and }$p_{s}$\textit{\ and }$q_{s}$\textit{\ be
defined by} $p_{s}=\left(  \left(  1-s\right)  2^{-1}+sp^{-1}\right)  ^{-1}$
\textit{and }$q_{s}=\left(  \left(  1-s\right)  2^{-1}+sq^{-1}\right)  ^{-1}%
$\textit{. Then for all }$s\in\left[  0,1\right]  $\textit{, the $\mu
$-deformed Segal-Bargmann transform }$B_{\mu}$\textit{\ is a bounded linear
map from }$L^{p_{s}}\left(  \mathbb{R},dg_{\mu}\right)  $\textit{\ to}
$L^{q_{s}}\left(  \mathbb{C}\times\mathbb{Z}_{2},d\nu_{\mu,\lambda}%
^{s}\right)  $\textit{, where}
\begin{align*}
d\nu_{\mu,\lambda}^{s}\left(  z,1\right)   &  =\left(  \kappa_{\lambda
,s}\left(  z,1\right)  \right)  ^{q_{s}}d\nu_{e,\mu}\left(  z\right)  ,\\
d\nu_{\mu,\lambda}^{s}\left(  z,-1\right)   &  =\left(  \kappa_{\lambda
,s}\left(  z,-1\right)  \right)  ^{q_{s}}d\nu_{o,\mu}\left(  z\right)  .
\end{align*}
\textit{\ } \textit{Moreover, for }$s\in\left[  0,1\right]  $ \textit{\ and}
$f\in{L^{p_{s}}\left(  \mathbb{R},dg_{\mu}\right)  }$\textit{\ we have that}
\begin{equation}
\left\Vert B_{\mu}f\right\Vert _{L^{q_{s}}\left(  \mathbb{C}\times
\mathbb{Z}_{2},d\nu_{\mu,\lambda}^{s}\right)  }\leq\left\Vert B_{\mu
}\right\Vert _{p\rightarrow q}^{s}\left\Vert f\right\Vert _{L^{p_{s}}\left(
\mathbb{R},dg_{\mu}\right)  }\text{.}\tag{6.3}%
\end{equation}

\bigskip

\textbf{Proof: \ }First let us note that for $s=0$ the measure $d\nu
_{\mu,\lambda}^{s}$ is simply $d\nu_{\mu}$, while for $s=1$ we have that
\begin{align*}
d\nu_{\mu,\lambda}^{1}\left(  z,1\right)   &  =\left(  \kappa_{\lambda
,1}\left(  z,1\right)  \right)  ^{q}d\nu_{e,\mu}\left(  z\right) \\
&  =\frac{\lambda^{\frac{2\mu+3}{2}}K_{\mu-\frac{1}{2}}\left(  \lambda
\left\vert z\right\vert ^{2}\right)  }{K_{\mu-\frac{1}{2}}\left(  \left\vert
z\right\vert ^{2}\right)  }\frac{2^{\frac{1}{2}-\mu}}{\pi\Gamma\left(
\mu+\frac{1}{2}\right)  }K_{\mu-\frac{1}{2}}\left(  \left\vert z\right\vert
^{2}\right)  \left\vert z\right\vert ^{2\mu+1}dxdy\\
&  =\lambda\frac{2^{\frac{1}{2}-\mu}}{\pi\Gamma\left(  \mu+\frac{1}{2}\right)
}K_{\mu-\frac{1}{2}}\left(  \lambda\left\vert z\right\vert ^{2}\right)
\left\vert \lambda^{\frac{1}{2}}z\right\vert ^{2\mu+1}dxdy\\
&  =d\nu_{\mu,\lambda}\left(  z,1\right)  .
\end{align*}

Similarly we have $d\nu_{\mu,\lambda}^{1}\left(  z,-1\right)  =d\nu
_{\mu,\lambda}\left(  z,-1\right)  $. That is, the measure $d\nu_{\mu,\lambda
}^{1}$ is $d\nu_{\mu,\lambda}$.

With the notation of Stein's theorem, we take $\left(  \Omega_{1},d\nu
_{1}\right)  =\left(  \mathbb{R},dg_{\mu}\right)  $ and $\left(  \Omega
_{2},d\nu_{2}\right)  =\left(  \mathbb{C}\times\mathbb{Z}_{2},d\nu_{\mu
}\right)  $. Take also $p_{0}=q_{0}=2$, $p_{1}=p$, $q_{1}=q$, $u_{0}%
,u_{1}:\mathbb{R}\rightarrow\left[  0,\infty\right)  $, $u_{0}\left(
t\right)  =u_{1}\left(  t\right)  \equiv1$, and $k_{0}:\mathbb{C}%
\times\mathbb{Z}_{2}\rightarrow\left[  0,\infty\right)  $, $k_{0}\left(
z,j\right)  \equiv1$, $j=-1,1$. Define $k_{1}:\mathbb{C}\times\mathbb{Z}%
_{2}\rightarrow\left[  0,\infty\right)  $ as $k_{1}:=\kappa_{\lambda,1}$,
where $\kappa_{\lambda,1}$ is defined in Lemma 6.1.

For $s\in\left[  0,1\right]  $, the function $u_{s}:\mathbb{R}\rightarrow
\left[  0,\infty\right)  $ in Stein's theorem is $u_{s}=u_{0}^{1-s}u_{1}^{s}=1
$, and the function $k_{s}:\mathbb{C}\times\mathbb{Z}_{2}\rightarrow\left[
0,\infty\right)  $ in Stein's theorem is $k_{s}=k_{0}^{1-s}k_{1}^{s}=k_{1}%
^{s}=\kappa_{\lambda,s}$, where $\kappa_{\lambda,s}$ is the function described
in Lemma 6.1.

Observe that
\begin{equation}
\left\Vert \left(  B_{\mu}f\right)  k_{s}\right\Vert _{L^{q_{s}}\left(
\mathbb{C}\times\mathbb{Z}_{2},d\nu_{\mu}\right)  }=\left\Vert B_{\mu
}f\right\Vert _{L^{q_{s}}\left(  \mathbb{C}\times\mathbb{Z}_{2},d\nu
_{\mu,\lambda}^{s}\right)  },\tag{6.4}%
\end{equation}
since
\begin{align*}
&  \left\Vert \left(  B_{\mu}f\right)  k_{s}\right\Vert _{L^{q_{s}}\left(
\mathbb{C}\times\mathbb{Z}_{2},d\nu_{\mu}\right)  }^{q_{s}}\\
&  =\int_{\mathbb{C}}\left\vert \left(  B_{e,\mu}f\right)  \left(  z\right)
\right\vert ^{q_{s}}\left(  \kappa_{\lambda,s}\left(  z,1\right)  \right)
^{q_{s}}d\nu_{e,\mu}(z)\\
&  +\int_{\mathbb{C}}\left\vert \left(  B_{o,\mu}f\right)  \left(  z\right)
\right\vert ^{q_{s}}\left(  \kappa_{\lambda,s}\left(  z,-1\right)  \right)
^{q_{s}}d\nu_{o,\mu}(z)\\
&  =\int_{\mathbb{C}}\left\vert \left(  B_{e,\mu}f\right)  \left(  z\right)
\right\vert ^{q_{s}}d\nu_{\mu,\lambda}^{s}\left(  z,1\right)  +\int
_{\mathbb{C}}\left\vert \left(  B_{o,\mu}f\right)  \left(  z\right)
\right\vert ^{q_{s}}d\nu_{\mu,\lambda}^{s}\left(  z,-1\right) \\
&  =\left\Vert B_{\mu}f\right\Vert _{L^{q_{s}}\left(  \mathbb{C}%
\times\mathbb{Z}_{2},d\nu_{\mu,\lambda}^{s}\right)  }^{q_{s}}.
\end{align*}

For $s=0$ we have
\[
\left\Vert \left(  B_{\mu}f\right)  k_{0}\right\Vert _{L^{2}\left(
\mathbb{C}\times\mathbb{Z}_{2},d\nu_{\mu}\right)  }=\left\Vert B_{\mu
}f\right\Vert _{L^{2}\left(  \mathbb{C}\times\mathbb{Z}_{2},d\nu_{\mu}\right)
}=\left\Vert f\right\Vert _{L^{2}\left(  \mathbb{R},dg_{\mu}\right)  }.
\]

Here the first equality is (6.4) and the second one is simply the fact that
the $\mu$-deformed Segal-Bargmann transform $B_{\mu}$ from $L^{2}\left(
\mathbb{R},dg_{\mu}\right)  $ to $L^{2}\left(  \mathbb{C}\times\mathbb{Z}%
_{2},d\nu_{\mu}\right)  $ is an isometry. That is, the hypothesis
(6.1)\textit{\ }of Stein's theorem is satisfied for $i=0$ with $A_{0}=1$.

For $s=1$ we have
\[
\left\Vert \left(  B_{\mu}f\right)  k_{1}\right\Vert _{L^{q}\left(
\mathbb{C}\times\mathbb{Z}_{2},d\nu_{\mu}\right)  }=\left\Vert B_{\mu
}f\right\Vert _{L^{q}\left(  \mathbb{C}\times\mathbb{Z}_{2},d\nu_{\mu,\lambda
}\right)  }\leq A_{1}\left\Vert f\right\Vert _{L^{p}\left(  \mathbb{R}%
,dg_{\mu}\right)  }.
\]

Here the first equality is again (6.4) and the second one is justified by the
fact that $B_{\mu}$ from $L^{p}\left(  \mathbb{R},dg_{\mu}\right)  $ to
$L^{q}\left(  \mathbb{C}\times\mathbb{Z}_{2},d\nu_{\mu,\lambda}\right)  $ is
bounded, by Theorem 4.1. Then, the hypothesis (6.1) of Stein's theorem is
satisfied for $i=1$ with $A_{1}=\left\Vert B_{\mu}\right\Vert _{p\rightarrow
q}$.

Thus, Stein's theorem allows us to conclude that the operator $B_{\mu}$ from
$L^{p_{s}}\left(  \mathbb{R},dg_{\mu}\right)  $ to $L^{q_{s}}\left(
\mathbb{C}\times\mathbb{Z}_{2},d\nu_{\mu,\lambda}^{s}\right)  $ is bounded and
that
\[
\left\Vert B_{\mu}f\right\Vert _{L^{q_{s}}\left(  \mathbb{C}\times
\mathbb{Z}_{2},d\nu_{\mu,\lambda}^{s}\right)  }\leq A_{1}^{s}\left\Vert
f\right\Vert _{L^{p_{s}}\left(  \mathbb{R},dg_{\mu}\right)  },
\]
as we wanted. \hfill\textbf{Q.E.D.}

\bigskip

The log-Sobolev inequality proved in [Snt1] involves the term $\left\langle
f,Nf\right\rangle _{L^{2}\left(  \mathbb{R},dg\right)  }$, called the
\textit{Dirichlet energy} (in the space $L^{2}\left(  \mathbb{R},dg\right)
$), which is the quadratic form associated to the number (or energy) operator
$N$. This operator is defined as $N=a^{\ast}a$, where $a^{\ast}$ and $a$ are
the creation and annihilation operators, respectively, acting in the ground
state representation $L^{2}\left(  \mathbb{R},dg\right)  $. The operators
$a^{\ast}$ and $a$ can be defined by their action on the elements $\zeta
_{n}\left(  t\right)  =2^{-\frac{n}{2}}\left(  n!\right)  ^{-\frac{1}{2}}%
H_{n}\left(  t\right)  $, $n=0,1,2,...$ (where $H_{n}\left(  t\right)  $ is
the $n$-th Hermite polynomial), which form an orthonormal basis of
$L^{2}\left(  \mathbb{R},dg\right)  $. The definitions are $a^{\ast}\zeta
_{n}=\left(  n+1\right)  ^{\frac{1}{2}}\zeta_{n+1}$ and $a\zeta_{n}=n
^{\frac{1}{2}}\zeta_{n-1}$, where $n=0,1,2,...$, and one defines $\zeta
_{-1}=0$. It turns out that $a^{\ast}$ is the adjoint of $a$ (with adequate
definitions of their domains, which we do not give here). Observe that
$N\zeta_{n}=n\zeta_{n}$, so $\zeta_{n}$ is an eigenvector of $N$ and $n$ is
the corresponding eigenvalue. This justifies the name \textquotedblleft number
operator\textquotedblright\ for $N$. Observe also that
\[
\left\langle f,Nf\right\rangle _{L^{2}\left(  \mathbb{R},dg\right)
}=\left\langle af,af\right\rangle _{L^{2}\left(  \mathbb{R},dg\right)
}=\left\|  af\right\|  _{L^{2}\left(  \mathbb{R},dg\right)  }^{2}= \frac{1}%
{2}\left\|  f^{\prime}\right\|  _{L^{2}\left(  \mathbb{R},dg\right)  }^{2}.
\]

That is, the Dirichlet energy is, up to a constant, the norm (in the space
$L^{2}\left(  \mathbb{R},dg\right)  $) of the derivative of the function $f$
(belonging to the domain of $N$), which is the Dirichlet form of $f$. Notice
that this is one of the ingredients of the Sobolev inequalities mentioned at
the beginning of this section.

The Segal-Bargmann transform $B:L^{2}\left(  \mathbb{R},dg\right)
\rightarrow\mathcal{B}^{2}$ intertwines the action of $a^{\ast}$ and $a$ for
the domain and codomain spaces, in the sense that $Ba^{\ast}B^{-1}$ and
$BaB^{-1}$ are the corresponding creation and annihilation operators in the
Segal-Bargmann space $\mathcal{B}^{2}$. We will continue denoting these
operators as $a^{\ast}$ and $a$ (acting on $\mathcal{B}^{2}$). It turns out
that $\left(  a^{\ast}f\right)  \left(  z\right)  =zf\left(  z\right)  $, and
$\left(  af\right)  \left(  z\right)  =f^{\prime}\left(  z\right)  $, where
$f^{\prime}$ is the complex derivative of the holomorphic function $f$.
Observe that since $B$ is unitary we have that $\left\langle f,Nf\right\rangle
_{L^{2}\left(  \mathbb{R},dg\right)  }=\left\langle Bf,\widetilde
{N}Bf\right\rangle _{\mathcal{B}^{2}},$ where $\widetilde{N}=BNB^{-1}$ is the
number operator in $\mathcal{B}^{2}$. That is, the Segal-Bargmann transform
$B$ preserves the Dirichlet energy (one says simply that ``$B$ preserves energy'').

For $\mu>-\frac{1}{2}$, the $\mu$-deformed generalizations of the results
above began to be considered in [Ros1], [Ros2] and [Marr], where the $\mu
$-deformed creation $a_{\mu}^{\ast}$ and annihilation $a_{\mu}$ operators in
the $\mu$-deformed quantum configuration space $L^{2}\left(  \mathbb{R}%
,\left\vert t\right\vert ^{2\mu}dt\right)  $ are defined. These definitions
are given in terms of the $\mu$-deformed position operator $\left(  Q_{\mu
}f\right)  \left(  t\right)  =tf\left(  t\right)  $ and the $\mu$-deformed
momentum operator $\left(  P_{\mu}f\right)  \left(  t\right)  =-i\left(
D_{\mu}f\right)  \left(  t\right)  $, where $\left(  D_{\mu}f\right)  \left(
t\right)  :=f^{\prime}\left(  t\right)  +\frac{\mu}{t}\left(  f\left(
t\right)  -f\left(  -t\right)  \right)  $. We mention in passing that $D_{\mu
}$, which is a $\mu$-deformation of the derivative, is a special case of a
more general class of operators called Dunkl operators (see [R\"{o}s]). The
definitions of $a_{\mu}^{\ast}$ and $a_{\mu}$ are $a_{\mu}^{\ast}=2^{-\frac
{1}{2}}\left(  Q_{\mu}-iP_{\mu}\right)  $ and $a_{\mu}=2^{-\frac{1}{2}}\left(
Q_{\mu}+iP_{\mu}\right)  $. The corresponding $\mu$-deformed creation and
annihilation operators in $L^{2}\left(  \mathbb{R},dg_{\mu}\right)  $ can be
defined by their action on the polynomials $\zeta_{n}^{\mu}\left(  t\right)
=2^{-\frac{n}{2}}\left(  n!\right)  ^{-1}\left(  \gamma_{\mu}(n)\right)
^{\frac{1}{2}}H_{n}^{\mu}\left(  t\right)  $, $n=0,1,2,...$, (where
$H_{n}^{\mu}\left(  t\right)  $ is the $\mu$-deformed Hermite polynomial of
degree $n$; see Section 3), which form an orthonormal basis of $L^{2}\left(
\mathbb{R},dg_{\mu}\right)  $. The definitions are $a_{\mu}^{\ast}\zeta
_{n}^{\mu}=\left(  n+1+2\mu\theta\left(  n+1\right)  \right)  ^{\frac{1}{2}%
}\zeta_{n+1}^{\mu}$ and $a_{\mu}\zeta_{n}^{\mu}=\left(  n+2\mu\theta\left(
n\right)  \right)  ^{\frac{1}{2}}\zeta_{n-1}^{\mu}$, where one defines
$\zeta_{-1}^{\mu}=0$. By considering the orthonormal basis $\left\{  \xi
_{n}^{\mu}\right\}  _{n=0}^{\infty}$ of the $\mu$-deformed Segal-Bargmann
space $\mathcal{B}_{\mu}^{2}$, one can define the Segal-Bargmann transform
$B_{\mu}:L^{2}\left(  \mathbb{R},dg_{\mu}\right)  \rightarrow\mathcal{B}_{\mu
}^{2}$ as $B\left(  \zeta_{n}^{\mu}\right)  =\xi_{n}^{\mu}$, $n=0,1,2,...$.
From this definition it is clear that $B_{\mu}$ is a unitary onto operator. It
is easy to see that the creation and annihilation operators on $\mathcal{B}%
_{\mu}^{2}$ are $\left(  a_{\mu}^{\ast}f\right)  \left(  z\right)  =zf\left(
z\right)  $ and $a_{\mu}f=\widetilde{D}_{\mu}f$, respectively. Here
$\widetilde{D}_{\mu}$ acts on holomorphic functions $f\left(  z\right)  $ as
$\widetilde{D}_{\mu}f\left(  z\right)  :=f^{\prime}\left(  z\right)
+\frac{\mu}{z}\left(  f\left(  z\right)  -f\left(  -z\right)  \right)  $,
where $f^{\prime}$ is the complex derivative of $f$. The $\mu$-deformed number
operator on $\mathcal{B}_{\mu}^{2}$ is $\widetilde{N}_{\mu}=a_{\mu}^{\ast
}a_{\mu}$, and one easily checks that $\widetilde{N}_{\mu}\xi_{n}^{\mu
}=\left(  n+2\mu\theta\left(  n\right)  \right)  \xi_{n}^{\mu}$, $n=0,1,2,...
$.

In [A-S.1] a $\mu$-deformed energy $E_{\mu}$ is introduced for functions
$f\in\mathcal{B}_{\mu}^{2}$, which appears as a term in a reverse log-Sobolev
inequality proved there. The definition is
\begin{equation}
E_{\mu}\left(  f\right)  =E_{e,\mu}\left(  f_{e}\right)  +E_{o,\mu}\left(
f_{o}\right)  ,\tag{6.5}%
\end{equation}
where $f_{e}$ and $f_{o}$ are the even and odd parts of $f$, respectively,
and
\begin{align*}
E_{e,\mu}\left(  f_{e}\right)   &  =\int_{\mathbb{C}}\left|  f_{e}\left(
z\right)  \right|  ^{2}\left|  z\right|  ^{2}d\nu_{e,\mu}(z),\\
E_{o,\mu}\left(  f_{o}\right)   &  =\int_{\mathbb{C}}\left|  f_{o}\left(
z\right)  \right|  ^{2}\left|  z\right|  ^{2}d\nu_{o,\mu}(z).
\end{align*}

When $\mu=0$ we have that $d\nu_{e,0}=d\nu_{o,0}=d\nu_{\text{Gauss}}$ and
(6.5) becomes
\begin{align*}
E_{0}\left(  f\right)   &  =\int_{\mathbb{C}}\left\vert f_{e}\left(  z\right)
\right\vert ^{2}\left\vert z\right\vert ^{2}d\nu_{\text{Gauss}}(z)+\int
_{\mathbb{C}}\left\vert f_{o}\left(  z\right)  \right\vert ^{2}\left\vert
z\right\vert ^{2}d\nu_{\text{Gauss}}(z)\\
&  =\int_{\mathbb{C}}\left\vert f\left(  z\right)  \right\vert ^{2}\left\vert
z\right\vert ^{2}d\nu_{\text{Gauss}}(z).
\end{align*}

(In the last equality we used that $zf_{e}\left(  z\right)  \in\mathcal{B}%
_{o}^{2}$, $zf_{o}\left(  z\right)  \in\mathcal{B}_{e}^{2}$, and that
$\mathcal{B}_{o}^{2}$ and $\mathcal{B}_{e}^{2}$ are orthogonal subspaces of
$\mathcal{B}^{2}$.)

In [Bar] it is proved that
\begin{equation}
\int_{\mathbb{C}}\left|  f\left(  z\right)  \right|  ^{2}\left|  z\right|
^{2}d\nu_{\text{Gauss}}\left(  z\right)  =\left\|  f\right\|  _{\mathcal{B}%
^{2}}^{2}+\left\langle f,\widetilde{N}f\right\rangle _{\mathcal{B}^{2}%
},\tag{6.6}%
\end{equation}
where $f\in\mathcal{B}^{2}$. This result (Bargmann identity) shows that in the
case $\mu=0$ the $\mu$-deformed energy defined above is related with the
Dirichlet energy $\left\langle f,\widetilde{N}f\right\rangle _{\mathcal{B}%
^{2}}$ for $f\in\mathcal{B}^{2}$. The $\mu$-deformed number operator $N_{\mu}
$ acting in $\mathcal{B}_{\mu}^{2}$ and its corresponding quadratic form
$\left\langle f,N_{\mu}f\right\rangle _{\mathcal{B}_{\mu}^{2}},$ which can be
identified as a $\mu$-deformed Dirichlet form, seem to have been introduced in
[A-S.1]. The relation of this Dirichlet energy and the $\mu$-deformed energy
$E_{\mu}\left(  f\right)  $ is studied in [A-S.2].

In the log-Sobolev inequality we will prove in this section there appears a
new mathematical object that it is natural to relate with the energy. We will
call it \textit{dilation energy}, and its definition is the following.

\bigskip

\textbf{Definition 6.1 }\textit{The dilation energy of an even function }%
$f\in\mathcal{B}_{e,\mu}^{2}$\textit{\ is defined by}%

\[
E_{e,\mu,\lambda}\left(  f\right)  =\int_{\mathbb{C}}\left\vert f\left(
z\right)  \right\vert ^{2}\log\left(  \frac{K_{\mu-\frac{1}{2}}\left(
\left\vert z\right\vert ^{2}\right)  }{K_{\mu-\frac{1}{2}}\left(
\lambda\left\vert z\right\vert ^{2}\right)  }\right)  d\nu_{e,\mu}(z).
\]

\textit{The dilation energy of an odd function }$f\in\mathcal{B}_{o,\mu}^{2}%
$\textit{\ is defined by}%

\[
E_{o,\mu,\lambda}\left(  f\right)  =\int_{\mathbb{C}}\left\vert f\left(
z\right)  \right\vert ^{2}\log\left(  \frac{K_{\mu+\frac{1}{2}}\left(
\left\vert z\right\vert ^{2}\right)  }{K_{\mu+\frac{1}{2}}\left(
\lambda\left\vert z\right\vert ^{2}\right)  }\right)  d\nu_{o,\mu}(z).
\]

\textit{The dilation energy of a function }$f\in\mathcal{B}_{\mu}^{2}%
$\textit{\ is defined by}%

\begin{equation}
E_{\mu,\lambda}\left(  f\right)  =E_{e,\mu,\lambda}\left(  f_{e}\right)
+E_{o,\mu,\lambda}\left(  f_{o}\right)  .\tag{6.7}%
\end{equation}

\bigskip

Observe that the fact that $\lambda\geq1$ and the decreasing property of
$K_{\alpha}\left(  x\right)  $ for $x\in\mathbb{R}^{+}$ imply that $\log
\frac{K_{\alpha}\left(  \left|  z\right|  ^{2}\right)  }{K_{\alpha}\left(
\lambda\left|  z\right|  ^{2}\right)  }\geq0$, so we have that $E_{\mu
,\lambda}\left(  f\right)  \geq0$.

When $\mu=0$ we can use (2.5) to obtain%
\begin{align*}
E_{0,\lambda}(f)  &  = \int_{\mathbb{C}}\left(  \left\vert f_{e}(z)
\right\vert ^{2}\!\log\!\left(  \! \frac{K_{-\frac{1}{2}}(\left\vert
z\right\vert ^{2})}{K_{-\frac{1}{2}}(\!\lambda\! \left\vert z\right\vert ^{2}
)}\!\right)  \!+\! \left\vert f_{o}(z) \right\vert ^{2}\!\log\!\left(  \!
\frac{K_{\frac{1}{2}}( \left\vert z\right\vert ^{2}) }{K_{\frac{1}{2}%
}(\!\lambda\! \left\vert z\right\vert ^{2}) }\!\right)  \right)
\!d\nu_{\text{Gauss}}(z)\\
&  =\int_{\mathbb{C}}\left(  \left\vert f_{e}\left(  z\right)  \right\vert
^{2}+\left\vert f_{o}\left(  z\right)  \right\vert ^{2}\right)  \log\left(
\tfrac{\left(  \frac{\pi}{2\left\vert z\right\vert ^{2}}\right)  ^{\frac{1}%
{2}}\exp\left(  -\left\vert z\right\vert ^{2}\right)  }{\left(  \frac{\pi
}{2\lambda\left\vert z\right\vert ^{2}}\right)  ^{\frac{1}{2}}\exp\left(
-\lambda\left\vert z\right\vert ^{2}\right)  }\right)  d\nu_{\text{Gauss}%
}(z)\\
&  =\int_{\mathbb{C}}\left(  \left\vert f_{e}\left(  z\right)  \right\vert
^{2}+\left\vert f_{o}\left(  z\right)  \right\vert ^{2}\right)  \left(
\log\lambda^{\frac{1}{2}}+\left(  \lambda-1\right)  \left\vert z\right\vert
^{2}\right)  d\nu_{\text{Gauss}}(z)\\
&  =\left(  \log\lambda^{\frac{1}{2}}\right)  \left\Vert f\right\Vert
_{\mathcal{B}^{2}}^{2}+\left(  \lambda-1\right)  \left(  \int_{\mathbb{C}%
}\left\vert f\left(  z\right)  \right\vert ^{2}\left\vert z\right\vert
^{2}d\nu_{\text{Gauss}}(z)\right)  .
\end{align*}

By using the Bargmann identity (6.6), we can write
\[
E_{0,\lambda}\left(  f\right)  =\left(  \log\lambda^{\frac{1}{2}}%
+\lambda-1\right)  \left\Vert f\right\Vert _{\mathcal{B}^{2}}^{2}+\left(
\lambda-1\right)  \left\langle f,\widetilde{N}f\right\rangle _{\mathcal{B}%
^{2}},
\]
which shows that, in the case $\mu=0$, the dilation energy $E_{\mu,\lambda
}\left(  f\right)  $ is related with the Dirichlet energy $\left\langle
f,\widetilde{N}f\right\rangle _{\mathcal{B}^{2}}$, where $f\in\mathcal{B}^{2}$.

In fact, for any $\mu>0$, the dilation energy $E_{\mu,\lambda}$ is related
with the $\mu$-deformed energy $E_{\mu}$, as we will see now.

We will use the following formula for $K_{\nu}\left(  x\right)  $, valid for
$x\in\mathbb{R}^{+}$ and $\nu>-\frac{1}{2}$. (See [Wat], p. 207.)
\begin{equation}
K_{\nu}(x)=\left(  \frac{\pi}{2x}\right)  ^{\frac{1}{2}}e^{-x}\left(
\sum_{k=0}^{n-1}\frac{\Gamma\left(  \nu\!+\!\frac{1}{2}\!+\!k\right)
}{k!\Gamma\left(  \nu\!+\!\frac{1}{2}\!-\!k\right)  \left(  2x\right)  ^{k}}+
\frac{\eta(x) \Gamma\left(  \nu\!+\!\frac{1}{2}\!+\!n\right)  }{n!\Gamma
\left(  \nu\!+\!\frac{1}{2}\!-\!n\right)  \left(  2x\right)  ^{n}}\right)
.\tag{6.8}%
\end{equation}

Here $\eta\left(  x\right)  $ is a function of $x$, $0\leq\eta\left(
x\right)  \leq1$, and the non-negative integer $n$ is chosen such that
$n-1<\nu-\frac{1}{2}\leq n$.

From (6.8) we obtain that
\[
\frac{K_{\mu-\frac{1}{2}}\left(  \left\vert z\right\vert ^{2}\right)  }%
{K_{\mu-\frac{1}{2}}\left(  \lambda\left\vert z\right\vert ^{2}\right)
}=\lambda^{\frac{1}{2}}\exp\left(  \left(  \lambda-1\right)  \left\vert
z\right\vert ^{2}\right)  S\left(  z,\mu,\lambda\right)  ,
\]
where for $z\ne0$
\[
S\left(  z,\mu,\lambda\right)  =\frac{\sum_{k=0}^{m-1}\frac{\Gamma\left(
\mu+k\right)  }{k!\Gamma\left(  \mu-k\right)  \left(  2\left|  z\right|
^{2}\right)  ^{k}}+\eta\left(  \left|  z\right|  ^{2}\right)  \frac
{\Gamma\left(  \mu+m\right)  }{m!\Gamma\left(  \mu-m\right)  \left(  2\left|
z\right|  ^{2}\right)  ^{m}}}{\sum_{k=0}^{m-1}\frac{\Gamma\left(
\mu+k\right)  }{k!\Gamma\left(  \mu-k\right)  \left(  2\lambda\left|
z\right|  ^{2}\right)  ^{k}}+\eta\left(  \lambda\left|  z\right|  ^{2}\right)
\frac{\Gamma\left(  \mu+m\right)  }{m!\Gamma\left(  \mu-m\right)  \left(
2\lambda\left|  z\right|  ^{2}\right)  ^{m}}},
\]
$\eta\left(  \left|  z\right|  ^{2}\right)  ,\eta\left(  \lambda\left|
z\right|  ^{2}\right)  \in\left[  0,1\right]  $, and $m\in\mathbb{N\cup
}\left\{  0\right\}  $ is such that $m<\mu\leq m+1$.

Similarly we have that
\[
\frac{K_{\mu+\frac{1}{2}}\left(  \left\vert z\right\vert ^{2}\right)  }%
{K_{\mu+\frac{1}{2}}\left(  \lambda\left\vert z\right\vert ^{2}\right)
}=\lambda^{\frac{1}{2}}\exp\left(  \left(  \lambda-1\right)  \left\vert
z\right\vert ^{2}\right)  T\left(  z,\mu,\lambda\right)  ,
\]
where for $z\ne0$
\[
T\left(  z,\mu,\lambda\right)  =\frac{\sum_{k=0}^{n-1}\frac{\Gamma\left(
\mu+1+k\right)  }{k!\Gamma\left(  \mu+1-k\right)  \left(  2\left|  z\right|
^{2}\right)  ^{k}}+\eta\left(  \left|  z\right|  ^{2}\right)  \frac
{\Gamma\left(  \mu+1+n\right)  }{n!\Gamma\left(  \mu+1-n\right)  \left(
2\left|  z\right|  ^{2}\right)  ^{n}}}{\sum_{k=0}^{n-1}\frac{\Gamma\left(
\mu+1+k\right)  }{k!\Gamma\left(  \mu+1-k\right)  \left(  2\lambda\left|
z\right|  ^{2}\right)  ^{k}}+\eta\left(  \lambda\left|  z\right|  ^{2}\right)
\frac{\Gamma\left(  \mu+1+n\right)  }{n!\Gamma\left(  \mu+1-n\right)  \left(
2\lambda\left|  z\right|  ^{2}\right)  ^{n}}},
\]
$\eta\left(  \left|  z\right|  ^{2}\right)  ,\eta\left(  \lambda\left|
z\right|  ^{2}\right)  \in\left[  0,1\right]  $, and $n\in\mathbb{N\cup
}\left\{  0\right\}  $ is such that $n-1<\mu\leq n.$

Thus, the dilation energy (6.7) can be written as
\begin{align*}
&  E_{\mu,\lambda}\left(  f\right) \\
&  =\int_{\mathbb{C}}\left\vert f_{e}\left(  z\right)  \right\vert ^{2}%
\log\left(  \lambda^{\frac{1}{2}}\exp\left(  \left(  \lambda-1\right)
\left\vert z\right\vert ^{2}\right)  S\left(  z,\mu,\lambda\right)  \right)
d\nu_{e,\mu}(z)\\
&  +\int_{\mathbb{C}}\left\vert f_{o}\left(  z\right)  \right\vert ^{2}%
\log\left(  \lambda^{\frac{1}{2}}\exp\left(  \left(  \lambda-1\right)
\left\vert z\right\vert ^{2}\right)  T\left(  z,\mu,\lambda\right)  \right)
d\nu_{o,\mu}(z)\\
&  =\left(  \log\lambda^{\frac{1}{2}}\right)  \left\Vert f\right\Vert
_{\mathcal{B}_{\mu}^{2}}^{2}\\
&  +\left(  \lambda-1\right)  \left(  \int_{\mathbb{C}}\left\vert f_{e}\left(
z\right)  \right\vert ^{2}\left\vert z\right\vert ^{2}d\nu_{e,\mu}%
(z)+\int_{\mathbb{C}}\left\vert f_{o}\left(  z\right)  \right\vert
^{2}\left\vert z\right\vert ^{2}d\nu_{o,\mu}(z)\right) \\
&  +\int_{\mathbb{C}}\left\vert f_{e}\left(  z\right)  \right\vert ^{2}\left(
\log S\left(  z,\mu,\lambda\right)  \right)  d\nu_{e,\mu}(z)\\
&  +\int_{\mathbb{C}}\left\vert f_{o}\left(  z\right)  \right\vert ^{2}\left(
\log T\left(  z,\mu,\lambda\right)  \right)  d\nu_{o,\mu}(z).
\end{align*}

That is, for any $\mu\geq0$ and $\lambda\geq1$ we have that the dilation
energy $E_{\mu,\lambda}\left(  f\right)  $ of a function $f\in\mathcal{B}%
_{\mu}^{2}$ is related with the $\mu$-deformed energy $E_{\mu}\left(
f\right)  $ by
\[
E_{\mu,\lambda}\left(  f\right)  =\left(  \log\lambda^{\frac{1}{2}}\right)
\left\Vert f\right\Vert _{\mathcal{B}_{\mu}^{2}}^{2}+\left(  \lambda-1\right)
E_{\mu}\left(  f\right)  +\rho\left(  \mu,\lambda,f\right)  ,
\]
where
\begin{align*}
&  \rho\left(  \mu,\lambda,f\right) \\
&  =\int_{\mathbb{C}}\left\vert f_{e}\left(  z\right)  \right\vert ^{2}\left(
\log S\left(  z,\mu,\lambda\right)  \right)  d\nu_{e,\mu}(z)+\int_{\mathbb{C}%
}\left\vert f_{o}\left(  z\right)  \right\vert ^{2}\left(  \log T\left(
z,\mu,\lambda\right)  \right)  d\nu_{o,\mu}(z).
\end{align*}

By examining the last factor in (6.8) we see that $S( z, \mu, \lambda) \to1 $
as $|z| \to\infty$. This also follows from (2.8). Similarly, $T( z, \mu,
\lambda) \to1 $ as $|z| \to\infty$. Moreover, $S( z, \mu, \lambda) \ne0 $ for
all $z \ne0$, since otherwise $K_{\mu- 1/2} ( |z|^{2} ) = 0 $, which is known
to be false. Similarly, $T( z, \mu, \lambda) \ne0 $ for all $z \ne0$. It is
then not hard to see that there exist constants $0 < A_{\mu,\lambda} <
B_{\mu,\lambda} $ such that
\[
A_{\mu,\lambda} ||f ||^{2}_{ \mathcal{B}^{2}_{\mu}} \le\rho( \mu, \lambda, f)
\le B_{\mu,\lambda} ||f ||^{2}_{ \mathcal{B}^{2}_{\mu}}%
\]
for all $f \in\mathcal{B}^{2}_{\mu}$. It follows for $\lambda> 1$ that the
quadratic forms $E_{\mu,\lambda} (f)$ and $E_{\mu}(f)$ in $f \in
\mathcal{B}^{2}_{\mu}$ are equivalent, modulo terms that are multiples of $||f
||^{2}_{ \mathcal{B}^{2}_{\mu}} $. Of course, for $\lambda= 1 $ we have
$E_{\mu,\lambda} (f) = 0 $ for all $f \in\mathcal{B}^{2}_{\mu}$.

Now we go to the main result of this section.

\bigskip

\textbf{Theorem 6.3 }(Logarithmic Sobolev Inequalities) \textit{Let }%
$p,q$\textit{\ be such that }
\[
1\leq q<2\lambda\qquad\text{\textit{and}\qquad}p>1+\frac{q}{2\lambda}.
\]
\textit{Let }$f\in L^{2+\zeta}\left(  \mathbb{R},dg_{\mu}\right)  $\textit{,
where }$\zeta>0$\textit{, be such that }$B_{\mu}f\in L^{2+\xi}\left(
\mathbb{C\times Z}_{2},d\nu_{\mu}\right)  $\textit{, where }$\xi>0$\textit{.
Then we have the logarithmic Sobolev inequality}
\begin{align}
&  \left(  2^{-1}-q^{-1}\right)  S_{L^{2}\left(  \mathbb{C\times Z}_{2}%
,d\nu_{\mu}\right)  }\left(  B_{\mu}f\right)  -\left(  2^{-1}-p^{-1}\right)
S_{L^{2}\left(  \mathbb{R},dg_{\mu}\right)  }\left(  f\right) \tag{6.9}\\
&  \leq\frac{1}{q}E_{\mu,\lambda}\left(  B_{\mu}f\right)  +\left(
\log\left\Vert B_{\mu}\right\Vert _{p\rightarrow q}-\frac{2\mu+3}{2q}%
\log\lambda\right)  \left\Vert f\right\Vert _{L^{2}\left(  \mathbb{R},dg_{\mu
}\right)  }^{2}.\nonumber
\end{align}

\bigskip

\textbf{Remark:} Consider the subspace $S$ of $L^{2} ( \mathbb{R} , dg_{\mu})
$ consisting of all $f$ such that $f \in L^{2+\zeta} ( \mathbb{R} , dg_{\mu})
$ for some $\zeta> 0$ and
$B_{\mu}f \in L^{2+\xi} ( \mathbb{C} \times\mathbb{Z}_{2} , d\nu_{\mu})$ for
some $\xi> 0$. Then $S$ is dense in $L^{2} ( \mathbb{R} , dg_{\mu})$. To see
this, first observe that the polynomials $\zeta_{n}^{\mu}$ are in
$L^{2+\alpha} ( \mathbb{R} , dg_{\mu})$ for every $\alpha> 0$, since the
density of the measure contains a Gaussian factor which dominates the
integrand near infinity. Now $B_{\mu}\zeta_{n}^{\mu}= \xi_{n}^{\mu}$ as we
already know. But $\xi_{n}^{\mu}$ is a monomial and so $\xi_{n}^{\mu}\in
L^{2+\beta} ( \mathbb{C} \times\mathbb{Z}_{2} , d\nu_{\mu})$ for every $\beta>
0$, since again the measure goes to zero fast enough to guarantee convergence
of the integral. Therefore, $\zeta_{n}^{\mu}\in S$ for every integer $n \ge0$.
But the set of finite linear combinations of the $\zeta_{n}^{\mu}$ forms a
subspace of $S$ which itself is dense in $L^{2} ( \mathbb{R} , dg_{\mu})$,
since the $\zeta_{n}^{\mu}$ are an orthonormal basis of $L^{2} ( \mathbb{R} ,
dg_{\mu})$. And this shows that $S$ is dense. Consequently, Theorems 5.2 and
6.3 hold for functions in a dense subspace, namely $S$, of $L^{2} ( \mathbb{R}
, dg_{\mu})$. We fully expect that the results of these two theorems hold for
all functions in $L^{2} ( \mathbb{R} , dg_{\mu})$.

\bigskip

\textbf{Proof of Theorem 6.3 : \ }We will use the inequality (6.3), which
combined with (6.4) tells us that for $s\in\left[  0,1\right]  $ we have that
\begin{equation}
\left\Vert \left(  B_{\mu}f\right)  k_{s}\right\Vert _{L^{q_{s}}\left(
\mathbb{C}\times\mathbb{Z}_{2},d\nu_{\mu}\right)  }\leq A^{s}\left\Vert
f\right\Vert _{L^{p_{s}}\left(  \mathbb{R},dg_{\mu}\right)  },\tag{6.10}%
\end{equation}
where $A=\left\Vert B_{\mu}\right\Vert _{p\rightarrow q}$ and $f$ is as in the
hypotheses of the theorem. Also $p_{s}$ and $q_{s}$ are as in Theorem 6.1.
Observe that when $s=0$, (6.10) becomes an equality, so we can differentiate
(6.10) at $s=0^{+}$ to obtain the new inequality
\[
\left.  \frac{d}{ds}\right\vert _{s=0^{+}}\left\Vert \left(  B_{\mu}f\right)
k_{s}\right\Vert _{L^{q_{s}}\left(  \mathbb{C}\times\mathbb{Z}_{2},d\nu_{\mu
}\right)  }\leq\left.  \frac{d}{ds}\right\vert _{s=0^{+}}\left(
A^{s}\left\Vert f\right\Vert _{L^{p_{s}}\left(  \mathbb{R},dg_{\mu}\right)
}\right)
\]
or
\begin{equation}
\left.  \frac{d}{ds}\right\vert _{s=0^{+}}\!\left\Vert \left(  B_{\mu
}f\right)  k_{s}\right\Vert _{L^{q_{s}}\left(  \mathbb{C}\times\mathbb{Z}%
_{2},d\nu_{\mu}\right)  }\leq\left(  \log A\right)  \!\left\Vert f\right\Vert
_{L^{2}\left(  \mathbb{R},dg_{\mu}\right)  }+\left.  \frac{d}{ds}\right\vert
_{s=0^{+}}\!\left\Vert f\right\Vert _{L^{p_{s}}\left(  \mathbb{R},dg_{\mu
}\right)  }.\tag{6.11}%
\end{equation}

The hypothesis on $f$ allows us to use Lemma 5.1 and obtain
\[
\left.  \frac{d}{ds}\right\vert _{s=0^{+}}\left\Vert f\right\Vert _{L^{p_{s}%
}\left(  \mathbb{R},dg_{\mu}\right)  }=\left(  2^{-1}-p^{-1}\right)
\left\Vert f\right\Vert _{L^{2}\left(  \mathbb{R},dg_{\mu}\right)  }%
^{-1}S_{L^{2}\left(  \mathbb{R},dg_{\mu}\right)  }\left(  f\right)  .
\]

The hypothesis on $B_{\mu}f$ and Lemma 6.1 imply $\left(  B_{\mu}f\right)
k_{s}\in L^{2+\xi}\left(  \mathbb{C\times Z}_{2},d\nu_{\mu}\right)  $ for
$s\in\left[  0,1\right]  $, and so we can also use Lemma 5.1 to obtain the
derivative of the left hand side of (6.11). Observe that in this case the
function $F$ of Lemma 5.1 is not a constant function. Thus, in this case
formula (5.3) gives us
\begin{align*}
&  \left.  \frac{d}{ds}\right\vert _{s=0^{+}}\left\Vert \left(  B_{\mu
}f\right)  k_{s}\right\Vert _{L^{q_{s}}\left(  \mathbb{C}\times\mathbb{Z}%
_{2},d\nu_{\mu}\right)  }\\
&  =\left\Vert B_{\mu}f\right\Vert _{L^{2}\left(  \mathbb{C\times Z}_{2}%
,d\nu_{\mu}\right)  }^{-1}\left(  2^{-1}-q^{-1}\right)  S_{L^{2}\left(
\mathbb{C\times Z}_{2},d\nu_{\mu}\right)  }\left(  B_{\mu}f\right) \\
&  +\left\Vert B_{\mu}f\right\Vert _{L^{2}\left(  \mathbb{C\times Z}_{2}%
,d\nu_{\mu}\right)  }^{-1}\operatorname{Re}\left\langle F^{\prime}\left(
0\right)  ,\left(  \operatorname*{sgn}B_{\mu}f\right)  \left\vert B_{\mu
}f\right\vert \right\rangle ,
\end{align*}
where the derivative $F^{\prime}\left(  0\right)  $ is
\[
F^{\prime}\left(  0\right)  =\left(  B_{\mu}f\right)  \left.  \frac{d}%
{ds}\right\vert _{s=0^{+}}k_{1}^{s}=\left(  B_{\mu}f\right)  \log k_{1},
\]
and so
\begin{align*}
\operatorname{Re}\left\langle F^{\prime}\left(  0\right)  ,\left(
\operatorname*{sgn}B_{\mu}f\right)  \left\vert B_{\mu}f\right\vert
\right\rangle  &  =\operatorname{Re}\int_{\mathbb{C}\times\mathbb{Z}_{2}%
}F^{\prime}\left(  0\right)  \overline{\left(  \operatorname*{sgn}\left(
B_{\mu}f\right)  \right)  }\left\vert B_{\mu}f\right\vert d\nu_{\mu}\\
&  =\operatorname{Re}\int_{\mathbb{C}\times\mathbb{Z}_{2}}\left(  \log
k_{1}\right)  \left(  B_{\mu}f\right)  \overline{B_{\mu}f}d\nu_{\mu}\\
&  =\int_{\mathbb{C}\times\mathbb{Z}_{2}}\left(  \log k_{1}\right)  \left\vert
B_{\mu}f\right\vert ^{2}d\nu_{\mu}.
\end{align*}

Explicitly we have that
\begin{align*}
&  \operatorname{Re}\left\langle F^{\prime}\left(  0\right)  ,\left(
\operatorname*{sgn}B_{\mu}f\right)  \left\vert B_{\mu}f\right\vert
\right\rangle \\
&  =\int_{\mathbb{C}}\log\left(  \frac{\lambda^{\frac{2\mu+3}{2}}K_{\mu
-\frac{1}{2}}\left(  \lambda\left\vert z\right\vert ^{2}\right)  }%
{K_{\mu-\frac{1}{2}}\left(  \left\vert z\right\vert ^{2}\right)  }\right)
^{q^{-1}}\left\vert \left(  B_{e,\mu}f\right)  \left(  z\right)  \right\vert
^{2}d\nu_{e,\mu}\left(  z\right) \\
&  +\int_{\mathbb{C}}\log\left(  \frac{\lambda^{\frac{2\mu+3}{2}}K_{\mu
+\frac{1}{2}}\left(  \lambda\left\vert z\right\vert ^{2}\right)  }%
{K_{\mu+\frac{1}{2}}\left(  \left\vert z\right\vert ^{2}\right)  }\right)
^{q^{-1}}\left\vert \left(  B_{o,\mu}f\right)  \left(  z\right)  \right\vert
^{2}d\nu_{o,\mu}\left(  z\right) \\
&  =\frac{2\mu+3}{2q}\left(  \log\lambda\right)  \left\Vert B_{\mu
}f\right\Vert _{L^{2}\left(  \mathbb{C\times Z}_{2},d\nu_{\mu}\right)  }%
^{2}-\frac{1}{q}E_{\mu,\lambda}\left(  B_{\mu}f\right)  .
\end{align*}

Thus we have that
\begin{align*}
&  \left.  \frac{d}{ds}\right\vert _{s=0^{+}}\left\Vert \left(  B_{\mu
}f\right)  k_{s}\right\Vert _{L^{q_{s}}\left(  \mathbb{C}\times\mathbb{Z}%
_{2},d\nu_{\mu}\right)  }\\
&  =\left\Vert B_{\mu}f\right\Vert _{L^{2}\left(  \mathbb{C\times Z}_{2}%
,d\nu_{\mu}\right)  }^{-1}\left(
\begin{array}
[c]{c}%
\left(  2^{-1}-q^{-1}\right)  S_{L^{2}\left(  \mathbb{C\times Z}_{2},d\nu
_{\mu}\right)  }\left(  B_{\mu}f\right) \\
\!\!+\frac{2\mu+3}{2q}\!\left(  \log\lambda\right)  \! \left\Vert B_{\mu
}f\right\Vert _{L^{2}\left(  \mathbb{C\times Z}_{2},d\nu_{\mu}\right)  }%
^{2}-\frac{1}{q}E_{\mu,\lambda}\!\left(  B_{\mu}f\right)
\end{array}
\!\!\right)  .
\end{align*}

So the inequality (6.11) becomes
\begin{align*}
&  \left\Vert B_{\mu}f\right\Vert _{L^{2}\left(  \mathbb{C\times Z}_{2}%
,d\nu_{\mu}\right)  }^{-1}\left(
\begin{array}
[c]{c}%
\left(  2^{-1}-q^{-1}\right)  S_{L^{2}\left(  \mathbb{C\times Z}_{2},d\nu
_{\mu}\right)  }\left(  B_{\mu}f\right) \\
\!\!+\frac{2\mu+3}{2q}\!\left(  \log\lambda\right)  \! \left\Vert B_{\mu
}f\right\Vert _{L^{2}\left(  \mathbb{C\times Z}_{2},d\nu_{\mu}\right)  }%
^{2}-\frac{1}{q}E_{\mu,\lambda}\!\left(  B_{\mu}f\right)
\end{array}
\!\!\right) \\
&  \leq\left(  \log A\right)  \left\Vert f\right\Vert _{L^{2}\left(
\mathbb{R},dg_{\mu}\right)  }+\left(  2^{-1}-p^{-1}\right)  \left\Vert
f\right\Vert _{L^{2}\left(  \mathbb{R},dg_{\mu}\right)  }^{-1}S_{L^{2}\left(
\mathbb{R},dg_{\mu}\right)  }\left(  f\right)  ,
\end{align*}
and finally, by using that $\left\Vert B_{\mu}f\right\Vert _{L^{2}\left(
\mathbb{C\times Z}_{2},d\nu_{\mu}\right)  }=\left\Vert f\right\Vert
_{L^{2}\left(  \mathbb{R},dg_{\mu}\right)  }$, we get
\begin{align*}
&  \left(  2^{-1}-q^{-1}\right)  S_{L^{2}\left(  \mathbb{C\times Z}_{2}%
,d\nu_{\mu}\right)  }\left(  B_{\mu}f\right)  -\left(  2^{-1}-p^{-1}\right)
S_{L^{2}\left(  \mathbb{R},dg_{\mu}\right)  }\left(  f\right) \\
&  \leq\frac{1}{q}E_{\mu,\lambda}\left(  B_{\mu}f\right)  +\left(
\log\left\Vert B_{\mu}\right\Vert _{p\rightarrow q}-\frac{2\mu+3}{2q}%
\log\lambda\right)  \left\Vert f\right\Vert _{L^{2}\left(  \mathbb{R},dg_{\mu
}\right)  }^{2},
\end{align*}
which is (6.9). \hfill\textbf{Q.E.D.} \nopagebreak

\bigskip

Observe that in the limiting case $\lambda=1$, we have that $E_{\mu,\lambda
}\left(  B_{\mu}f\right)  =0$, and then the log-Sobolev inequality (6.9)
becomes
\begin{align*}
&  \left(  \frac{1}{2}-\frac{1}{q}\right)  S_{L^{2}\left(  \mathbb{C\times
Z}_{2},d\nu_{\mu}\right)  }\left(  B_{\mu}f\right) \\
&  \leq\left(  \frac{1}{2}-\frac{1}{p}\right)  S_{L^{2}\left(  \mathbb{R}%
,dg_{\mu}\right)  }\left(  f\right)  +\left(  \log\left\Vert B_{\mu
}\right\Vert _{p\rightarrow q}\right)  \left\Vert f\right\Vert _{L^{2}\left(
\mathbb{R},dg_{\mu}\right)  }^{2},
\end{align*}
which is the Hirschman inequality (5.4) we proved in the previous section.

In the case $\mu=0$ the inequality (6.9) becomes
\begin{align*}
&  \left(  2^{-1}-q^{-1}\right)  S_{L^{2}\left(  \mathbb{C},d\nu
_{\text{Gauss}}\right)  }\left(  Bf\right)  -\left(  2^{-1}-p^{-1}\right)
S_{L^{2}\left(  \mathbb{R},dg\right)  }\left(  f\right) \\
&  \leq\frac{1}{q}E_{0,\lambda}\left(  Bf\right)  +\left(  \log\left\Vert
B\right\Vert _{p\rightarrow q}-\frac{3}{2q}\log\lambda\right)  \left\Vert
f\right\Vert _{L^{2}\left(  \mathbb{R},dg\right)  }^{2}%
\end{align*}
or
\begin{align*}
&  \left(  2^{-1}-q^{-1}\right)  S_{L^{2}\left(  \mathbb{C},d\nu
_{\text{Gauss}}\right)  }\left(  Bf\right)  -\left(  2^{-1}-p^{-1}\right)
S_{L^{2}\left(  \mathbb{R},dg\right)  }\left(  f\right) \\
&  \leq\frac{1}{q}\left(  \left(  \log\lambda^{\frac{1}{2}}+\lambda-1\right)
\left\Vert Bf\right\Vert _{\mathcal{B}^{2}}^{2}+\left(  \lambda-1\right)
\left\langle Bf,\widetilde{N}Bf\right\rangle _{\mathcal{B}^{2}}\right) \\
&  +\left(  \log\left\Vert B\right\Vert _{p\rightarrow q}-\frac{3}{2q}%
\log\lambda\right)  \left\Vert f\right\Vert _{L^{2}\left(  \mathbb{R}%
,dg\right)  }^{2}.
\end{align*}

By using that $\left\Vert Bf\right\Vert _{\mathcal{B}^{2}}=\left\Vert
f\right\Vert _{L^{2}\left(  \mathbb{R},dg\right)  }^{2}$ and $\left\langle
Bf,\widetilde{N}Bf\right\rangle _{\mathcal{B}^{2}}=\left\langle
f,Nf\right\rangle _{L^{2}\left(  \mathbb{R},dg\right)  }$, we can write the
last expression as
\begin{align*}
&  \left(  2^{-1}-q^{-1}\right)  S_{L^{2}\left(  \mathbb{C},d\nu
_{\text{Gauss}}\right)  }\left(  Bf\right)  -\left(  2^{-1}-p^{-1}\right)
S_{L^{2}\left(  \mathbb{R},dg\right)  }\left(  f\right) \\
&  \leq\left(  -\frac{1}{q}\log\lambda+\frac{\lambda-1}{q}+\log\left\Vert
B\right\Vert _{p\rightarrow q}\right)  \left\Vert f\right\Vert _{L^{2}\left(
\mathbb{R},dg\right)  }^{2}+\frac{\lambda-1}{q}\left\langle f,Nf\right\rangle
_{L^{2}\left(  \mathbb{R},dg\right)  },
\end{align*}
which is the log-Sobolev inequality in [Snt1], up to some identifications in
the coefficients of the terms of the right hand side (for example, the weight
$a$ that appears in [Snt1] can be identified with $\lambda-1$).

\section{Concluding remarks}

\bigskip

In this section we present some of the lines along which this work can be continued.

(1) The $\mu$-deformed theory presented in [Ros1], [Ros2], and [Marr] is valid
for $\mu>-\frac{1}{2}$. Nevertheless, the inequality (2.2) was proved only for
non-negative values of $\mu$, and this inequality is fundamental in the proof
of the Theorem 4.1, and then in the proofs of results of the remaining
sections. We leave as open questions if these results (Sections 4, 5 and 6)
are also valid for $-\frac{1}{2}<\mu<0$.

(2) Theorem 4.1 establishes that if $p\in\left(  1,\infty\right]  $,
$q\in\left[  1,\infty\right)  $, and $\lambda>\frac{1}{2}$ are such that the
inequalities $p>1+\frac{q}{2\lambda}$ and $1\leq q<2\lambda$ hold, then the
$\mu$-deformed Segal-Bargmann transform $B_{\mu}$ is a bounded operator from
$L^{p}\left(  \mathbb{R},dg_{\mu}\right)  $ to $\mathcal{B}_{\mu,\lambda}^{q}%
$. For $p$, $q$, and $\lambda$ not satisfying the above mentioned inequalities
we know little about the boundedness of $B_{\mu}$. We suspect that if either
of the inequalities $q>2\lambda$ or $p<1+\frac{q}{2\lambda}$ holds, then
$B_{\mu}$ is not bounded (for the corresponding values of $p$, $q$ and
$\lambda$), since this is the case when $\mu=0$ and $\lambda=1$ (see Corollary
7.2 in [Snt1]), but in the general situation we consider in this work this
remains as an open question.

\bigskip

\section*{Acknowledgments}

\bigskip

The first author wishes to thank firstly CIMAT and the Mittag-Leffler
Institute (Djursholm, Sweden) for supporting his attendance in the Fall
Program \textit{Partial Differential Equations and Spectral Theory} at the
Mittag-Leffler Institute (October-December 2002) and secondly the Universidad
Panamericana (Mexico City) for giving him the support for being a full-time
doctoral student at CIMAT from 2001 to 2004. The second author thanks Larry
Thomas for many useful comments over the course of the years. Both authors
thank Shirley Bromberg, Pavel Naumkin, Roberto Quezada Batalla and Carlos
Villegas-Blas for valuable comments.

\end{document}